\documentclass[aps,prc,showpacs,twocolumn]{revtex4}
\usepackage{graphicx}
\usepackage{dcolumn}
\usepackage{bm}
\usepackage{longtable}
\usepackage{hyperref}
\usepackage{natbib}
\usepackage{amsmath}
\usepackage{multirow}
\usepackage{color}
\begin{document}
\title{The $0\nu\beta\beta$-decay nuclear matrix element for light and heavy neutrino mass mechanisms from deformed QRPA cacluations for $^{76}$Ge, $^{82}$Se, $^{130}$Te, $^{136}$Xe and $^{150}$Nd with isospin restoration.}
\author{Dong-Liang Fang$^{a}$ and Amand Faessler$^b$ and Fedor \v{S}imkovic$^{c,d}$}
\affiliation{$^a$College of Physics, Jilin University, Changchun, Jilin 130012, People's Republic of China}
\affiliation{$^b$Institute of Theoretical Physics, University of Tuebingen, D-72076 Tuebingen, Germany}
\affiliation{$^c$INR, 141980 Dubna, Moscow Region, Russia}
\affiliation{$^d$Comenius University, Physics Department, SK-842 15 Bratislava, Slovakia}
\begin{abstract}
In this work, with restored isospin symmetry, we evaluated the neutrinoless double beta decay nuclear matrix elements for $^{76}$Ge, $^{82}$Se, $^{130}$Te, $^{136}$Xe and $^{150}$Nd for both the light and heavy neutrino mass mechanisms using the deformed QRPA approach with realistic forces. We give detailed decompositions of the nuclear matrix elements over different intermediate states and nucleon pairs, and discuss how these decompositions are affected by the model space truncations. Compared to the spherical calculations, our results show reductions from $30\%$ to about $60\%$ of the nuclear matrix elements for the calculated isotopes mainly due to the presence of BCS overlap factor between the initial and final ground states. The comparison between different nucleon-nucleon forces  with corresponding Short-Range-Correlations (src) shows,  that the choice of the NN force gives roughly $20\%$ deviations for light exchange neutrino mechanism and much larger deviations for the heavy neutrino exchange mechanism.
\end{abstract}
\maketitle
\section{Introduction}
As the fundamental blocks of the Standard Model, neutrinos have been surrounded by mysteries since their discoveries. It is known that neutrinos have masses from the oscillation experiments, but it is still unclear how the masses are generated. Since the right-handed neutrinos are missing in Standard Model, the Yukawa-coupling of Fermions with Higgs bosons responsible for the mass generation of charged fermions are not available for neutrinos. So one naturally seeks to enlarge the gauge symmetry in order to include the right handed weak gauge bosons as well as the right-handed neutrinos. In addition one needs also  new Higgs bosons, which can  break the new symmetry and give mass to the neutrinos and the new gauge bosons. On the other hand the huge mass hierarchy between neutrinos and charged fermions also needs a satisfying answer. A quite promising solution to above questions called the See-Saw model (For a review see\cite{Moh05}) has been proposed with several variants. These models include besides the normal Dirac mass terms of Yukawa coupling, also Majorana mass terms. The mixing between the right- and left-hand neutrinos yields  very small mass eigenvalues for neutrinos and breaks the lepton number symmetry. With the introduction of the Majorana terms, new phenomena  will emerge as a consequence of lepton number violation, one of which is the so-called neutrinoless double beta decay ($0\nu\beta\beta$).

Double beta decay is a very rare decay originating from the mass staggering due to  nuclear pairing.  It has a half-life much longer than normal nuclear decays. It can occur in the Standard Model scenario, where an even-even nucleus decays to a neighboring even-even nucleus with the same mass number but with two more protons, and emits two electrons and two anti-electron-neutrinos, provided that the two even-even nuclei have a large enough mass difference. This is the two-neutrino double beta decay ($2\nu\beta\beta$). Since its discovery in the 1980's \cite{EHM87}, more than a dozen nuclei have been experimentally confirmed to decay with this mode (For a recent compilation see\cite{Bar15}). Today, more efforts are devoted to search for a more interesting mode beyond the Standard Model, the one without neutrino emission. As we have stated above, if neutrinos are Majorana particles, the emitted neutrinos can be reabsorbed in another weak vertex of the double beta decay. This should produce a small  peak at the upper end of the two-electron spectrum. This is the so-called neutrinoless double beta decay ($0\nu\beta\beta$). 
In the See-Saw theory, the small Standard Model neutrino mass comes from the existence of heavy right-handed neutrino. Thus there exist two kinds of neutrinos one with small masses (light) and another with large masses (heavy) mediating the process. Generally, the light neutrino mechanism will be the dominant one, but still very rare. With specific light neutrino mass eigenvalues and mixing angles, this mechanism could be suppressed. In this case, the dominant mechanism of $0\nu\beta\beta$ will be mediated by the heavy neutrino, therefore, calculation of $0\nu\beta\beta$ with heavy neutrino mechanism is also necessary.

For nuclear theorists, the most difficult part of the calculation is to understand the nuclear transition of the $\beta\beta$ process and this requires certain nuclear many-body approaches. Various methods have been adopted for evaluation of this process, which could be in general divided into two categories by how one deals with the intermediate states: one category is with the so-called closure approximation, which eliminates the intermediate states, such as the Shell Model \cite{CNP99}, the Projected Hatree-Fock-Boglyubov (HFB)  method \cite{CCR08}, the Interacting Boson Model (IBM) \cite{BKI13} and Density Functional Theory (DFT) of the relativistic \cite{SYR14} and the  non-relativistic versions \cite{RM10}. Approaches  without closure approximation are mainly Quasi-particle Random Phase Approximation (QRPA) \cite{PSV96,RFS06,SFM09,SRFV13,ME13,HS15} and partly also the Shell Model (see \cite{SHB14}). The advantage of non-closure methods are, that they can deal with both modes of the double beta decay, since for $2\nu\beta\beta$ the nuclear matrix elements  depend sensitively on the energies of the intermediate states. The $2\nu\beta\beta$ can then act as a test for the quality  of the many-body theory.

There are different versions of QRPA according to the choice of the mean field and residual interactions. QRPA with realistic forces have been used for double beta decay for decades \cite{PSV96,SRFV13,HS15}. These calculations could well reproduce the $2\nu\beta\beta$ NME's and give pretty good predictions of $0\nu\beta\beta$ NME's. There are also QRPA calculations with Skyrme forces for $2\nu\beta\beta$ \cite{MAS08} and $0\nu\beta\beta$ \cite{ME13} as well. The treatment  of deformation in QRPA for double beta decay can be traced back to more than a decade ago \cite{SPF03,ASM04}. Afterwards also  realistic forces \cite{YRF08} were used. Deformed calculation with Skyrme forces found \cite{ME13}, that deformations  play a role  even for slightly deformed nuclei. Therefore, deformed calculation for less deformed nuclei can give a much better understanding of NME's and can serve as cross check of the reliability of spherical calculations.

This work is arranged as follows: in section II we briefly introduce the formalism, then present the results of matrix elements of several nuclei in Section III and details of the matrix element structure for $^{150}$Nd in Section IV followed by the conclusion.

\section{Formalisms and methods}
The $0\nu\beta\beta$ half-life with the neutrino mass mechanism can be written as \cite{SPV99}:
\begin{eqnarray}
[T_{1/2}^{0\nu}]^{-1}=G^{01}|\langle m_\nu \rangle M_{l}^{0\nu}+\eta_N M_{h}^{0\nu}|^2
\end{eqnarray}
Where $G^{01}$ is the phase space factor for the emitted electrons and $M_{i}^{0\nu}$ (The subscripts $i\ = \ l$ stands for the light neutrino mechanism and $i \ = \ h$ for the heavy neutrino exchange.) is the matrix element presenting the nuclear transition of this process. $\langle m_\nu\rangle$ and $\eta_N$ are defined in the way as functions of PNMS (Pontecorvo-Maki-Nakagawa-Sakata) mixing matrix  and neutrino masses \cite{SPV99}: $\langle m_\nu \rangle=|\sum_{l} U_{el}^2m_l|$ and $\eta_N=|\sum_{h} U^2_{eh}\frac{m_p}{M_h}|$, here U's are PNMS matrix elements, $m_l$ and $M_h$ are mass eigenvalues for light and heavy neutrinos, respectively and $m_p$ is the proton mass. With the introduction of induced currents, the $0\nu\beta\beta$ NME's can be divided in three parts:
\begin{eqnarray}
M_i^{0\nu}=-\frac{M^{0\nu}_{F,i}}{g_A^2}+M^{0\nu}_{GT,i}+M^{0\nu}_{T,i}
\end{eqnarray} 
For each part, the detailed expression has been derived in \cite{SPV99} with the general form:
\begin{eqnarray}
M^{0\nu}_{K,i}=\langle f| H_{K,i}(r) \mathcal{O}_K |i\rangle
\end{eqnarray} 
Where the operators are $\mathcal{O}_F=1$, $\mathcal{O}_{GT}={\bf \sigma_1\cdot \sigma_2}$ and $\mathcal{O}_T=3{\bf (\sigma_1\cdot\hat{r})(\sigma_2\cdot\hat{r})-\sigma_1\cdot\sigma_2}$ for Fermi, Gamow-Teller(GT) and Tensor parts respectively. $H_{K,i}(r)$ is the so-called neutrino potential which is the integration of the neutrino propagator over the neutrino momentum with the form\cite{SFR07}:
\begin{eqnarray}
H_{K,l}(r)&=&\frac{2}{\pi g_A^2} \frac{R}{r} \int_0^{\infty} \frac{\sin(qr)}{q+E_{J}^m-(E_{g.s.}^i+E_{g.s.}^f)/2} \nonumber\\
&\times&h_{K}(q^2)  dq\nonumber \\
H_{K,h}(r)&=&\frac{1}{m_p m_e}\frac{2}{\pi g_A^2} \frac{R}{r} \int_0^{\infty} \sin(qr)h_{K}(q^2) q dq
\end{eqnarray}
Here $R=1.2 \cdot A^{1/3}$ fm is the nuclear radius. $E_J^m$ is the energy of m-th. excited state of intermediate odd-odd nuclei and $E_{g.s.}^{i(f)}$ are the energies of the ground state of the initial (final) even-even nuclei.
 $h_K(q^2)$ is the respective form factors given in \cite{SPV99}.

In this work  we use the deformed Quasi-Particle Random Phase Approximation (QRPA) with realistic forces introduced by \cite{YRF08}for the nuclear many-body method, where the G-matrix is obtained  by the Brueckner equation. The G-matrix is used as residual force for pairing and the pn-QRPA phonons. 

The QRPA calculation starts with single particle wave functions solved from a Coulomb corrected Woods-Saxon potential \cite{YRF08}. The wave function with axial symmetry can be expanded into the spherical harmonic basis as \cite{SPF03}: 
\begin{eqnarray}
|\tau \Omega_{\tau} \rangle = B_{\tau\eta} | \eta \Omega_\tau \rangle
\end{eqnarray}
where $|\eta\Omega_\tau \rangle=C_{l m 1/2 s}^{j \Omega_\tau} | n_r l m\rangle |1/2, s\rangle $ is the spherical Harmonic basis with a finite angular momentum. 

With the above expression we can transform the calculations of one- and two- body matrix elements from integrations of deformed wave functions and operators to sums of the spherical matrix elements instead \cite{YRF08,FFR11}. The residual or pairing interaction matrix elements can then be expressed as decompositions over spherical G-matrix elements \cite{YRF08,FFR11}:
\begin{eqnarray}
V_{\tau_1\tau_2\tau_3\tau_4}&=&
\sum_{J}\sum_{\eta_{\tau_1}\eta_{\tau_2}}\sum_{\eta_{\tau_3}\eta_{\tau_4}} F_{\tau_1\eta_{\tau_1}\tau_2 \eta_{\tau_2}}^{JK}F_{\tau_3\eta_{\tau_3}\tau_4 \eta_{\tau_4}}^{JK} \nonumber \\
&\times& G(\eta_{\tau_1}\eta_{\tau_2}\eta_{\tau_3}\eta_{\tau_4},J)
\end{eqnarray}
Here $\tau_i$ represents either proton or neutron. And the transformation coefficients are defined as 
 \\   \\ 
 \hspace{1cm}  $F_{\tau_1\eta_{\tau_1}\tau_2 \eta_{\tau_2}}^{JK}\equiv B_{\tau_1 \eta_{\tau_1}}B_{\tau_1 \eta_{\tau_1}} (-1)^{j_{\tau_2}-\Omega_{\tau_2}} C_{j_{\tau_1}\Omega_{\tau_1}j_{\tau_2}\Omega_{\tau_2}}^{JK}$.
\\   \\   
With above interactions, we can solve the BCS equations. However, proper renormalization of the pairing strength is needed to reproduce the experimental pairing gaps.  To do this, we multiply the pairing interaction matrix elements with factors $g^{pair}_{p(n)}$.

With the solution of the BCS equations, we can further derive the proton-neutron(pn-) QRPA equations:
\begin{eqnarray}
\left( \begin{array}{cc}
A(K^\pi) & B(K^\pi) \\ -B(K^\pi) & -A(K^\pi)
\end{array} \right)
\left( \begin{array}{c}
X^{K^\pi} \\ Y^{K^\pi}
\end{array} \right)=\omega \left( \begin{array}{c}
X^{K^\pi} \\ Y^{K^\pi}
\end{array} \right)
\end{eqnarray}
Where:
\begin{eqnarray}
A_{pn,p'n'}({K^\pi})&=&\delta_{pp'}\delta_{nn'} (E_p+E_n) \nonumber \\
&-&2 g_{ph}(u_pv_nu_{p'}v_{n'}+v_{p}u_{n}v_{p'}u_{n'})V_{pnp'n'} \nonumber \\
&-&2 (u_pu_nu_{p'}u_{n'}+v_{p}v_{n}v_{p'}v_{n'}) \nonumber \\
&\times&(g_{pp}^{T=1}V^{T=1}_{p\tilde{n}p'\tilde{n}'}+g_{pp}^{T=0}V^{T=0}_{p\tilde{n}p'\tilde{n}'})  \\
B_{pn,p'n'}({K^\pi})&=&-2 g_{ph}(u_pv_nv_{p'}u_{n'}+v_{p}u_{n}u_{p'}v_{n'})V_{pnp'n'} \nonumber \\
&+&2 (u_pu_nv_{p'}v_{n'}+v_{p}v_{n}u_{p'}u_{n'}) \nonumber \\
&\times&(g_{pp}^{T=1}V^{T=1}_{p\tilde{n}p'\tilde{n}'}+g_{pp}^{T=0}V^{T=0}_{p\tilde{n}p'\tilde{n}'}) 
\end{eqnarray}
Here $g_{ph}$ and $g_{pp}$ are renormalized strengths of the residual interaction. In the particle-particle (pp) channel, we split the interactions into isoscalar (T=0) and isovector (T=1) parts with separate renormalization parameters as in \cite{RF11,SRFV13} for isospin symmetry restoration.

With the solutions of the QRPA equations, we can perform the calculations of double beta decay matrix elements. The NME's can be expressed in a general form as:
\begin{eqnarray}
M^{\beta\beta}_I&=&\sum_{K^\pi m_i m_f} \sum_{pnp'n'} \langle pn | \mathcal{O}_{I}^{\beta\beta} |p'n'\rangle \nonumber \\
 &\times&{}_f\langle 0| c_p^\dagger c_n | K^\pi m_f\rangle \langle K^\pi m_f|K^\pi m_i \rangle \langle K^\pi m_i| c_{p'}^\dagger c_{n'} | 0\rangle_i \nonumber \\
\end{eqnarray}
%\end{widetext}
Here $\beta\beta$ could be either $0\nu$ or $2\nu$. The nuclear transition matrix elements in QRPA calculations are:
\begin{eqnarray}
\langle K^\pi m_i | c_p^\dagger c_n | 0\rangle_i&=&X^{ m_f}_{pn,K^\pi}u_p v_n+Y^{ m_i}_{pn,K^\pi}v_p u_n \nonumber \\
{}_f\langle 0| c_p^\dagger c_n | K^\pi m_f\rangle&=&X^{ m_f}_{pn,K^\pi}v_p u_n+Y^{ m_f}_{pn,K^\pi}u_p v_n \nonumber \\
\langle K^\pi m_f|K^\pi m_i \rangle&=&\sum_{pnp'n'}\mathcal{R}_{pn,p'n'} {}_f\langle 0 | 0\rangle_i \nonumber \\
&\times&(X^{m_f}_{pn,K^\pi}X^{m_i}_{p'n',K^\pi}-Y^{m_f}_{pn,K^\pi}Y^{m_i}_{p'n',K^\pi}) \nonumber 
\end{eqnarray}
Where $\mathcal{R}_{pn,p'n'}$ is defined in \cite{YRF08} and ${}_f\langle 0 | 0\rangle_i$ is the BCS overlap factor denoting the overlaps between the parent and daughter nuclei expressed in \cite{SPF03}:
\begin{eqnarray}
&&{}_f\langle 0 | 0 \rangle_i \nonumber \\
&=&\prod_{k=1}^{N_{\Omega}} u_{k}^{(f)} \prod_{l=1}^{N_{\Omega}} u_{l}^{(i)} \nonumber \\
&+&\sum_{m_1,n_1=1}^{N_\Omega} v_{m_1}^{(f)} v_{n_1}^{(i)} (D^{(1)}(m_1;n_1))^2\prod_{k=1}^{N_{\Omega}(m_1)} u_{k}^{(f)} \prod_{l=1}^{N_{\Omega}(n_1)} u_{l}^{(i)} \nonumber \\
&+&\sum_{m_1,m_2,n_1,n_2=1}^{N_\Omega} v_{m_1}^{(f)} v_{m_2}^{(f)}v_{n_1}^{(i)} v_{n_2}^{(i)}(D^{(1)}(m_1,m_2;n_1,n_2))^2\nonumber \\
&\times&\prod_{k=1}^{N_{\Omega}(m_1,m_2)} u_{k}^{(f)} \prod_{l=1}^{N_{\Omega}(n_1,n_2)} u_{l}^{(i)} +\ldots 
\end{eqnarray}
Here $|0\rangle_{i(f)}$ are BCS vacua for initial and final nuclei respectively. $N_\Omega$ denotes the total number of single particle levels of the model space, $\prod_{k=1}^{N_{\Omega}(m_1,m_2)}$ means that the sum runs over the values from $1$ to $N_\Omega$ except $m_1$ and $m_2$. $D^r{(1)}(m_1,\ldots, m_r;n_1,\ldots,n_r)$ denotes the determinant of a matrix of rank $r$ constructed of the elements of the unitary matrix of the transformation between the initial and final single particle states with row indices $m_1,\ldots,m_r$ and column indices $n_1,\ldots,n_r$.  

For $M^{2\nu}_F$, the intermediate states are have only  $K^\pi=0^+$, while $K^\pi$ is summed over $0^+$ and $\pm 1^+$ for $M^{2\nu}_{GT}$, and all the $K^\pi$'s for $M^{0\nu}$. Due to  the axial symmetry of the wave-functions, there are degeneracies between the normal states and their time reversed states ($K^\pi$ and $-K^\pi$ for $K > 0$).  The single particle two-body matrix elements are calculated following the methods of interaction matrix element calculation:
\begin{eqnarray}
\langle pn | \mathcal{O}_F^{2\nu} |p'n' \rangle &=& \sum_{J\ge K} F_{p\eta_pn\eta_n}^{JK}F_{p'\eta_{p'}n'\eta_{n'}}^{JK} 
\frac{\langle p||\mathcal{I} ||n \rangle  \langle p'|| \mathcal{I} ||n' \rangle}{\bar{\omega}} \nonumber \\
\langle pn | \mathcal{O}_{GT}^{2\nu} |p'n' \rangle &=& \sum_{J\ge K} F_{p\eta_pn\eta_n}^{JK}F_{p'\eta_{p'}n'\eta_{n'}}^{JK} 
\frac{\langle p ||\sigma||n \rangle  \langle p' ||\sigma ||n' \rangle}{3\bar{\omega}} \nonumber \\
\langle pn | \mathcal{O}_I^{0\nu} |p'n' \rangle &=& \sum_{J\ge K} F_{p\eta_pn\eta_n}^{JK}F_{p'\eta_{p'}n'\eta_{n'}}^{JK} 
\langle p n||H_{I}(r)\mathcal{O}_I|p'n' \rangle  \nonumber
\end{eqnarray}
The energy denominator are defined as $\bar{\omega}\equiv\frac{E_{m_i}+E_{m_f}}{2}-E_{d}$ with $E_d=\frac{E_{g.s.}^i+E_{g.s.}^f}{2}$. In deformed calculations a closure energy $\bar{\omega}=7MeV+E_{g.s.,m}-E_d$ is used for $0\nu\beta\beta$, where $E_{g.s.,m}$ is the energy of the ground state of the intermediate odd-odd nucleus. The matrix elements $\langle p n||H_{I}(r)\mathcal{O}_I|p'n' \rangle$ in the spherical system are expressed in \cite{SPV99}. To account the hard repulsive core of the nuclear force, one usually rewrite the neutrino potential in the form $f(r)H_{K}(r)f(r)$, where $f(r)$ is the so-called short-range-correlation (src) function. In this work, we use the src consistent to the realistic NN forces use in a form $f(r)=1-c e^{-ar^2}(1-b r^2)$ with $a$ $b$ and $c$ detailed in \cite{SFM09} for different versions of short range correlations.

\section{Nuclear matrix elements for several nuclei}

\begin{table*}[htp]
\caption{The deformation parameters and pairing strengths for five $\beta\beta$-decaying nuclei as well as their products. We also present in this table the BCS overlaps between the initial and final nuclei as well as the particle-particle interaction strength in the T=1 and T=0 channels. }
\begin{center}
\renewcommand{\arraystretch}{1.5}
\begin{tabular}{|c|ccccccc|ccccccc|}
\toprule
&\multicolumn{7}{c|}{AV18}&\multicolumn{7}{c|}{CD Bonn}\\
 &$\beta_2$ &  $_i\langle 0|0 \rangle_f$&  $g^{pair}_{p}$ &  $g^{pair}_n$ &  $g_{pp}^{T=1}$ &  $g_{pp}^{T=0}(1.0)$ &  $g_{pp}^{T=0} (1.27)$&  $\beta_2$ &  $_i\langle 0|0 \rangle_f$&  $g^{pair}_{p}$ &  $g^{pair}_n$ &  $g_{pp}^{T=1}$ &  $g_{pp}^{T=0}(1.0) $ &  $g_{pp}^{T=0}(1.27) $ \\ 
  \hline
  $^{76}$Ge 
  & 0.24 &\multirow{2}{0.8cm}{0.72}&1.07 & 1.12 &\multirow{2}{0.8cm}{1.24}&\multirow{2}{0.8cm}{0.80}  & \multirow{2}{0.8cm}{0.85}
  & 0.24 &\multirow{2}{0.8cm}{0.73}&0.97 & 1.02 &\multirow{2}{0.8cm}{1.13}&\multirow{2}{0.8cm}{0.72}  & \multirow{2}{0.8cm}{0.77} \\
    $^{76}$Se 
  & 0.28 & 					      &1.22 & 1.18 &  & &
  & 0.28 & 					      & 1.11 & 1.07&  &  & \\
  \hline
    $^{82}$Se 
  & 0.16 &\multirow{2}{0.8cm}{0.71}  &0.94 & 1.21 &\multirow{2}{0.8cm}{1.21}&\multirow{2}{0.8cm}{0.78}  & \multirow{2}{0.8cm}{0.83}
  & 0.16 &\multirow{2}{0.8cm}{0.71}  &0.85 & 1.10 &\multirow{2}{0.8cm}{1.09}&\multirow{2}{0.8cm}{0.69}  & \multirow{2}{0.8cm}{0.75} \\
    $^{82}$Kr 
    &0.18 &						& 1.13 & 1.22 & & &
    &0.18 &						& 1.01 & 1.10 & & & \\
    \hline
    $^{130}$Te 
  & 0.12 &\multirow{2}{0.8cm}{0.73}  &1.02 & 1.07 &\multirow{2}{0.8cm}{1.14}&\multirow{2}{0.8cm}{0.77}  & \multirow{2}{0.8cm}{0.79}
  & 0.12 &\multirow{2}{0.8cm}{0.73}  &0.93 & 0.97 &\multirow{2}{0.8cm}{1.04}&\multirow{2}{0.8cm}{0.69}  & \multirow{2}{0.8cm}{0.71} \\
    $^{130}$Xe 
    &0.16 &						& 1.07 & 1.10 & & &
    &0.16 &						& 0.97 & 0.99 & & & \\
\hline
    $^{136}$Xe 
  & 0.08 &\multirow{2}{0.8cm}{0.43}  &0.91 & - &\multirow{2}{0.8cm}{1.10}&\multirow{2}{0.8cm}{0.65}  & \multirow{2}{0.8cm}{0.71}
  & 0.08 &\multirow{2}{0.8cm}{0.39}  &0.80 & - &\multirow{2}{0.8cm}{0.95}&\multirow{2}{0.8cm}{0.55}  & \multirow{2}{0.8cm}{0.60} \\
    $^{136}$Ba 
    &0.11 &						& 1.00 & 1.10 & & &
    &0.11 &						& 0.88 & 0.89 & & & \\
%\hline
%    $^{136}$Xe 
%  & 0.0 &\multirow{2}{0.8cm}{0.73}  &0.92 & - &\multirow{2}{0.8cm}{1.10}&\multirow{2}{0.8cm}{0.79}  & \multirow{2}{0.8cm}{0.80}
%  & 0.0 &\multirow{2}{0.8cm}{0.76}  &0.81 & - &\multirow{2}{0.8cm}{0.95}&\multirow{2}{0.8cm}{0.67}  & \multirow{2}{0.8cm}{0.68} \\
%    $^{136}$Ba 
%    &0.11 &						& 1.00 & 1.10 & & &
%    &0.11 &						& 0.88 & 0.89 & & & \\
    \hline
    $^{150}$Nd 
  & 0.24 &\multirow{2}{0.8cm}{0.51 }  &1.03 & 1.14 &\multirow{2}{0.8cm}{1.16 }&\multirow{2}{0.8cm}{ 0.81}  & \multirow{2}{0.8cm}{ 0.85}
  & 0.24 &\multirow{2}{0.8cm}{0.52}  &0.94 & 1.03 &\multirow{2}{0.8cm}{1.06}&\multirow{2}{0.8cm}{0.74}  & \multirow{2}{0.8cm}{0.77} \\
    $^{150}$Sm 
    &0.15 &						& 1.04 & 1.16 & & &
    &0.15 &						& 0.95 & 1.04 & & & \\
  \hline
\botrule
\end{tabular}
\end{center}
\vspace{-0.5cm}
\label{par}
\end{table*}

%BCS issue
In this section, we present the nuclear matrix elements for the $0\nu\beta\beta$-decay of both light and heavy exchanged-neutrinos for the five nuclei $^{76}$Ge, $^{82}$Se, $^{130}$Te, $^{136}$Xe and $^{150}$Nd. 

At first we present the parameters we use in our calculations. For the single particle energies and wave functions, we use the deformed Wood-Saxon potential with Coulomb corrections introduced in \cite{YRF08}, where the quadrupole deformation parameters $\beta_2$ are fitted from B(E2) data  in \cite{RNT01} using a procedure from \cite{FFR11}. For the two lighter nuclei, $^{76}$Ge and $^{82}$Se, we adopt a model space with 7 major shells for $N=0-6$, while for the other three isotopes, we use a larger model space with 8 major shells $N=0-7$. In this work, we use realistic G-matrix elements for both pairing and residual interactions. For the sake of comparison, we adopt two different realistic NN interactions: Argonne-V18 (AV18) \cite{WSS95} and the Charge Dependent Bonn (CD-Bonn) \cite{Mac01} interaction. For the pairing interactions, one introduces renormalization pairing strength parameters  denoted by $g_p^{pair}$ and $g_n^{pair}$ for protons and neutrons, respectively. These strengths are fixed by reproducing the pairing gaps obtained from the five-point formula \cite{AWT03}. Of these nuclei, $^{136}$Xe has a  magic neutron number. Including the deformation we could obtain a  BCS solution reproducing the pairing gap of neutrons. But this solution gives a pretty smooth Fermi surface and breaks the magicity of the neutrons. Thus  we follow the treatment in \cite{SRFV13} and switch off the neutron pairing for this nucleus in the calculation. These pairing parameters  $g_{pair}$'s are presented in Table \ref{par}. We find that different pairing strengths are needed for the two interactions. Generally, we need larger strengths for AV18 to get a correct pairing behavior. The so fitted pairing strength parameters yield results close to one, this means the interactions we use are close to the bare G-matrix elements. We find  at most $20\%$ deviations from bare G-matrix elements for both AV18 and the CD-Bonn potentials. The different pairing interactions lead however  to similar BCS solutions. This can be verified from the BCS overlaps factors between initial and final ground states $\ _{i}\langle 0 | 0\rangle_{f}$, they differ at most at the last digits as we can see from Table \ref{par}. 
As stated in \cite{SPF03}, these factors are mostly affected by the difference of the deformations of initial and final nuclei than the absolute  magnitude of the deformations themselves. In the current work most nuclei have differences for the deformation of the initial and final nuclei  in $\beta_2$ from 0.02 to 0.06 and the BCS overlap factor is 
around 0.7. But for $^{150}$Nd, the difference of the deformations is as large as  0.09, this leads to a smaller overlap factor around 0.5. We also have an exception here, for $^{136}$Xe, with  $0.03$ difference of the deformation, a small BCS overlap of 0.4 between the initial and final ground states is observed. This is mostly due to the different Fermi surfaces of initial non-paired and final paired neutrons. In general, we have two kinds of nuclei here, one with moderately larger  BCS overlaps and one with smaller BCS overlaps, this will affect the NME's and fitted parameters 
of $g_{pp}^{T=0}$ as we shall see.

%% gpp issue
To enforce the isospin symmetry conservation, we follow the treatment from \cite{RF11,SRFV13} by separating the fitting of T=0 (isoscalar) and T=1 (isovector) particle-particle (pp) residual interactions. Following the treatment in\cite{SRFV13}, $g_{pp}^{T=1}$ is fixed by the condition $M_F^{2\nu}=0$.  The fitted values of $g_{pp}$'s are presented in Table.\ref{par}, for the $T=1$ channel, as proven in \cite{RF11}, in order to restore the isospin symmetry, $g_{pp}^{T=1}$ should approximately equal to $g^{pair}$'s, this conclusion holds for deformed cases \cite{FFS15} in our previous calculation, and now for more nuclei we find that either for the AV18 or CD-Bonn potential, fitted $g_{pp}^{T=1}$'s are approximately equal to the average of the pairing strength of the initial (final) neutron and proton within the numerical accuracy. While compared to the spherical calculation \cite{SRFV13}, where one uses a similar Woods-Saxon potential but with H.O. wave functions and smaller model space, both fitted $g_{pair}$'s and $g_{pp}^{T=1}$ of the current calculations are relatively larger.

Meanwhile, the strength in the isoscalar channel, $g_{pp}^{T=0}$ is fixed by the experimental $2\nu\beta\beta$ nuclear matrix elements \cite{Bar15}. We have in our calculations two sets of $g_{pp}^{T=0}$ values corresponding to two cases of axial vector coupling constants $g_A$, one bare $g_A=g_{A0}\ = \ 1.27$ and another quenched $g_A=0.75 \cdot g_{A0}$. The strength in the  T=0 channel $g_{pp}^{T=0}$ is much smaller than that in T=1 channel for the two NN forces. We generally need a larger renormalization strength for AV18 than for CD Bonn. The introduction of BCS overlap factors doesn't  reduce strongly the values of $g_{pp}^{T=0}$,  if we compare the fitted values in the deformed case  to the spherical calculations in \cite{SRFV13}, where this overlap factor is not included. The reason for this is the sharp drop of NME's as $g_{pp}^{T=0}$ increases \cite{YRF08}, therefore, changes of NME's leads to small changes of $g_{pp}^{T=0}$. We see that the fitted $g_{pp}^{T=0}$'s are around $0.8$ for AV18 and $0.7$ for CD-Bonn, with the exemption of $^{136}$Xe, which has a very small BCS overlap. The BCS overlap factor for $^{136}$Xe may be too small and one needs perhaps a better pairing theory to simulate this overlap factor for nuclei close to magic numbers.

\begin{table*}[htp]
\caption{$0\nu\beta\beta$ matrix elements for the light neutrino mechanisms for the five isotopes. We present the results for two NN potentials (see text), and for each NN force two cases: with and without quenching of $g_A$. In the second column, "a" denotes the case without "src" (short range correlations)  and "b" with "src"  calculated with self-consistent Br\"uckner methods\cite{SFM09} for each  NN potential. $M'^{0\nu}$ is the total matrix element, see text.}
\begin{center}
\renewcommand{\arraystretch}{1.5}
\begin{tabular}{|c|c|cccc|cccc|cccc|cccc|}
\toprule
& &\multicolumn{8}{c|}{AV18}&\multicolumn{8}{c|}{CD Bonn}\\
\hline
& &\multicolumn{4}{c|}{$g_A=g_{A0}$}&\multicolumn{4}{c|}{$g_A=0.75g_{A0}$} &\multicolumn{4}{c|}{$g_A=g_{A0}$}&\multicolumn{4}{c|}{$g_A=0.75g_{A0}$}\\
 &  &  $M_{F}^{0\nu}$ &$M_{GT}^{0\nu}$& $M_{T}^{0\nu}$ & $M'^{0\nu}_l$
      &  $M_{F,l}^{0\nu}$ &$M_{GT,l}^{0\nu}$& $M_{T,l}^{0\nu}$ & $M'^{0\nu}_l$ 
      &  $M_{F}^{0\nu}$ &$M_{GT}^{0\nu}$& $M_{T}^{0\nu}$ & $M'^{0\nu}_l$
      &  $M_{F,l}^{0\nu}$ &$M_{GT,l}^{0\nu}$& $M_{T,l}^{0\nu}$ & $M'^{0\nu}_l$\\ 
  \hline
  $^{76}$Ge$\rightarrow ^{76}$Se 
             & a  & -1.09 & 3.11 & -0.44  & 3.34 & -1.09 & 3.94 & -0.46 & 2.63
                    & -1.10 & 2.99 & -0.40 & 3.27 & -1.09 & 3.90 & -0.42 &  2.64\\
             & b  & -1.06 & 2.92 & -0.45  &3.12 &  -1.06 & 3.70 & -0.47 & 2.48
                    & -1.15 & 3.09 & -0.41  & 3.40 & -1.15 & 4.00 & -0.43 & 2.72\\
  \hline
  $^{82}$Se$\rightarrow ^{82}$Kr 
  & a  & -1.00 & 2.86 & -0.41  & 3.07  & -1.00 & 3.61 & -0.43 &2.41
           & -1.00 & 2.76 & -0.37 & 3.01 & -1.00 & 3.58 & -0.42 & 2.41\\
  & b & -0.98 & 2.68 & -0.42  & 2.86 & -0.97 & 3.39 & -0.38 &2.26
            & -1.05 & 2.85 & -0.38  & 3.13 & -1.05 & 3.67 & -0.39 &2.49\\
  \hline
   $^{130}$Te$\rightarrow ^{130}$Xe
 & a  & -1.17 & 2.95 & -0.52  &3.16  & -1.16 & 3.37 & -0.55 &2.31
          & -1.15 & 2.85 & -0.46 &3.10  & -1.15 & 3.29 & -0.49 &2.29\\
& b & -1.13 & 2.73 & -0.53  & 2.90 & -1.13 & 3.11 & -0.56 & 2.13
          & -1.21 & 2.95 & -0.47  &3.22  & -1.21 & 3.38 & -0.50 &2.37\\
  \hline
  $^{136}$Xe$\rightarrow ^{136}$Ba 
  & a  & -0.37 & 1.12 & -0.17  & 1.18 & -0.37 & 1.39 & -0.17 & 0.91
           & -0.33 & 1.05 & -0.13 & 1.12 & -0.33 & 1.29 & -0.14 & 0.85\\
 &b & -0.36 & 1.06 & -0.17  & 1.11 & -0.36 & 1.31 & -0.17 &0.86
           & -0.35 & 1.10 & -0.14  & 1.18 & -0.35 & 1.33 & -0.14 & 0.89\\
%  \hline
%  $^{136}$Xe$\rightarrow ^{136}$Ba 
%  & a  & -1.31 & 2.90 & -0.51  &  & -1.31 & 3.23 & -0.54 & 
%           & -1.36 & 3.19 & -0.47 &  & -1.36 & 3.53 & -0.50 &\\
% &b & -1.27 & 2.67 & -0.52  &  & -1.27 & 2.95 & -0.55 &
%           & -1.42 & 3.33 & -0.48  &  & -1.43 & 3.66 & -0.51 &\\
  \hline
  $^{150}$Nd$\rightarrow ^{150}$Sm 
 & a & -1.35 & 2.98 & -0.53  & 3.28& -1.35 & 3.54 & -0.56 & 2.52
           & -1.36 & 2.89 & -0.45 & 3.28 & -1.37 & 3.45 & -0.52 & 2.50\\
 & b & -1.32 & 2.74 & -0.55  & 3.01 & -1.31 & 3.26 & -0.57 & 2.33
           & -1.43 & 3.00 & -0.46  & 3.43 & -1.43 & 3.55 & -0.53 &2.59\\
 \hline
\botrule
\end{tabular}
\end{center}
\vspace{-0.5cm}
\label{lnt}
\end{table*}

The matrix elements are presented in Table \ref{lnt} for the light and in table \ref{hnt} for the heavy neutrino case. As we have seen, the fitted $g_{pp}$'s which correctly reproduce $M_F^{2\nu}$ and $M_{GT}^{2\nu}$ are quite different for the two forces, AV18 and CD-Bonn. Nevertheless, these differences don't lead to large deviations for the  $0\nu\beta\beta$ NME's. This means that a set of parameters, which give the same $2\nu\beta\beta$ NME's, will give basically the same $0\nu\beta\beta$ NME's irrespective of the chosen realistic NN interactions. This is surprising, since one expects, that different forces will give different $0\nu\beta\beta$ NME's, and calculations show no obvious correlations  between $0\nu\beta\beta$ and $2\nu\beta\beta$ NME's. Thus one would not expect, that wave functions, which  lead to the same $2\nu\beta\beta$ NME's do necessarily lead also to the same $0\nu\beta\beta$ NME's. 

We start with the analysis of the light neutrino mass mechanism as discussed in our previous work \cite{FFS15}. The new parametrization, which partially restores the isospin symmetry in QRPA calculations, reduces the Fermi part $M^{0\nu}_F$ while the GT part remains unaffected. As a consequence, the new parameters will bring down the magnitude of the ratio $\chi_F=M^{0\nu}_F/M^{0\nu}_{GT}$ close to shell model predictions: $\chi_F=-1/3$ \cite{SRFV13}. In the deformed calculations, values of this factor is oscillating around $-1/3$ for both the quenching and non-quenching cases as one can see.

% inclusion of tensor
For the light neutrino mechanism, we now include also the tensor part from higher order currents, which reduce  to zero, if neutrino momenta are small. With a realistic nucleon form factor, we find, that tensor parts give reductions of about $15\%$ to the GT matrix elements $M^{0\nu}_{GT}$. Since $M_F\approx -1/3 M_{GT}$ this would give a reduction of about $10\%$ for the total matrix elements.  This differs from Shell Model calculations, where they found negligible contributions from the tensor part (See for example \cite{SH14}). On the other hand, we find that the Tensor part is much more sensitive to the NN  potentials, but since it's small in magnitude, this doesn't affect the above conclusions, that overall $0\nu\beta\beta$ NME's are not sensitive to the NN potentials.

% with or without quenching
The quenching of axial-vector coupling constant $g_A$ is always a problem in nuclear physics (For a recent review, see \cite{LMR17,Suh17}), and for $0\nu\beta\beta$, if quenching is included, we would in general observe a decrease of the total matrix elements \cite{SRFV13}. This originates from the fact that $0\nu\beta\beta$ and $2\nu\beta\beta$ NME's have a different $g_{pp}$ dependence. At first, we should be aware of that quenching does not affect the Fermi parts of the NME's in the new parametrization, since in our approach, $M_{F}^{2\nu}$ is always zero. So $g_{pp}^{T=1}$ is not affected by quenching neither, as a consequence, $M_F^{0\nu}$ is independent of quenching too. This can be seen in table \ref{lnt}. Thus, when we discuss the quenching effect, we consider only $M_{GT}$ and $M_{T}$. If $M^{0\nu}_{GT}$ and $M_{GT}^{2\nu}$ has the same dependence on $g_{pp}^{T=0}$, namely $M_{GT}^{2\nu}(g_{pp1})/M_{GT}^{0\nu}(g_{pp1})=M_{GT}^{2\nu}(g_{pp2})/M_{GT}^{0\nu}(g_{pp2})$, we can get $M_{GT,g_A}^{2(0)\nu}=(g_A/g_{A0})^2 M_{GT,g_{A0}}^{2(0)\nu}$. We then obtain the same total NME's $M'^{0\nu}$ with or without quenching. 
If $M^{0\nu}_{GT}$ changes much more drastic than $M^{2\nu}_{GT}$ as $g_{pp}^{T=0}$ changes, we will have larger total neutrinoless NME's. But in our calculations we find the opposite:  $M^{0\nu}_{GT}$ or $M^{0\nu}_{GT}+M^{0\nu}_T$ change much slower than $M^{2\nu}_{GT}$. When $M^{2\nu}_{GT}$ changed by $(1/0.75)^2-1\sim 80\%$, the change of $M^{0\nu}_{GT}$ is just about $30\%$, therefore we observe about $20\%\sim 30\%$ reductions of the total $0\nu\beta\beta$ NME's when quenching is included. Such a behavior of less drastic changes for $0\nu\beta\beta$ NME's derives from the fact, that $0\nu\beta\beta$ NME receives contributions from all intermediate $K^\pi$ states, while $2\nu\beta\beta$ only from $K^\pi=0^+,\pm1^+$. Our calculations show, that $0\nu\beta\beta$ and $2\nu\beta\beta$ NME's are only sensitive to $g_{pp}^{T=0}$ for $K^\pi=0^+,\pm1^+$ intermediate states. The sensitivity of other intermediate states are otherwise small\cite{Fan11}. This explains the different sensitivity and consequently the reduction of $M^{0\nu}_{GT}$ by quenching. The same may be applied for the  Tensor part. But on one hand the $K^\pi=0^+,\pm1^+$ intermediate states  contribute even less to the total $M_{T}$, and on the other hand $M_T$ contribute too little to the overall NME's. Thus the quenching effect is small for the Tensor parts. A thorough discussion of quenching effect on the NMEs and its connection to $g_{pp}$ is presented in \cite{Suh17PRC}. The origin of possible quenching of the axial-vector coupling constant is still unknown. Recently, it was shown that if quenching is due to two-body currents it has only small effect on $0\nu\beta\beta$ unlike it is in the case of $2\nu\beta\beta$\cite{MGS11,ESV14}. 

% with or without src
The src (short range correlation)  is another important issue in the actual calculation of $0\nu\beta\beta$. Early calculations use the Jastrow src's, which gives a relatively large more than $10\%$ reductions to the final results \cite{FFR11}. Modern self-consistent src's behave much milder \cite{SFM09,FFR11}. In this work, we follow the self-consistent treatment in \cite{SFM09}. For each of the two NN interactions used here, we get a general conclusion as in most earlier publications, that the src for CD-Bonn are much milder, while for Argonne V18 src seems to give larger corrections. In general, CD-Bonn src gives enhancement to the total NME for about several percent and the opposite for Argonne src which reduces the NME's by several percent. This leads to an about $10\%$ difference  for the overall $0\nu\beta\beta$ NME's $M'^{0\nu}$ from the two potentials Argonne V18 and CD-Bonn.

\begin{table}[htp]
\caption{Comparison of total $0\nu\beta\beta$ NME's in \cite{SRFV13} and current work without BCS overlap factors. For each force, we have there columns, the first column presents the results of spherical calculation, the second column are deformed calculations with $\beta_2=0$ for both initial and final nuclei and BCS overlap factors ${}_f\langle 0 | 0\rangle_i=1$ and the third column the deformed results divided by the BCS overlap factors.}
\begin{center}
\begin{tabular}{|c|ccc|ccc|}
\hline
		& \multicolumn{3}{c|}{AV-18} 	& \multicolumn{3}{c|}{CD-Bonn} \\
		& sph\cite{SRFV13}& def I & def II & sph\cite{SRFV13}& def I & def II \\
\hline
$^{76}$Ge &5.16 & 5.00 & 4.33 &5.57 & 5.45 & 4.66\\
\hline
$^{82}$Se & 4.64&	 4.71 & 4.03 & 5.02 & 5.18 & 4.41\\
\hline
$^{130}$Te & 3.89 & 3.88 & 3.97 & 4.37 & 4.37 & 4.41\\
\hline
$^{136}$Xe & 2.18 & 2.19 & 2.58 & 2.46 & 2.67 & 3.03 \\
\hline			
\end{tabular}
\end{center}
\label{sphdef}
\end{table}

%comparison with spherical results
For the five nuclei studied, four of which we have also spherical results \cite{SRFV13}, but not  for $^{150}$Nd, which is strongly deformed. For these four nuclei, we find that the difference between the two calculations mainly originates from the BCS overlap factors not included in the spherical approach \cite{SRFV13}. If we divide the  results of this work including deformations  by the BCS overlap factors, the results are comparable with the spherical calculations. % This is due to the fact that these four isotopes have small deformations, which do not change the single particle and quasi-particle levels largely. 
This can be seen from Table.\ref{sphdef}, these results show that the deformed calculations under the spherical limit basically reproduce the spherical calculations in \cite{SRFV13}, a negligible deviations of several percents are observed for these two cases. And in deformed calculations, if the absolute deformation is small($^{130}$Te case), the main difference of the deformed and spherical results comes from the BCS overlap factors, if the deformation increases, we may observe deviations beyond the BCS overlap factors. 
Therefore, compared with spherical calculations, the deformed calculations give reductions partly from the BCS overlaps between the initial and final ground states. And this reduction is about $30\%$ for $^{76}$Ge, $^{82}$Se and $^{130}$Te, 
and $60\%$ for $^{136}$Xe as explained above. This conclusion can be generalized to other nuclei calculated in \cite{SRFV13}.  If the neutron or proton number is not magic (as for  $^{96}$Zr or $^{100}$Mo), we would get a reduction of $20\sim 30\%$ to the calculated NME's, otherwise a larger reduction would be expected (for example $^{116}$Cd or $^{124}$Sn). If the absolute deformations are large for initial and final nuclei, we could observe a further decrease. This decrease is somehow nucleus dependent, in our calculation, it is large for $^{76}$Ge and $^{82}$Se, but small for $^{150}$Nd. Therefore, although under the spherical limit\cite{FFR11}, $0\nu\beta\beta$ NME for $^{150}$Nd is slightly larger than other nuclei, when the deformation is included, the NME divided by BCS overlap factor is much larger than that of $^{76}$Ge and $^{82}$Se, therefore although $^{150}$Nd has a small BCS overlap factor, it has total $0\nu\beta\beta$ NME comparable to that of $^{76}$Ge and $^{82}$Se.

We find for the LNM (Light Neutrino exchange Mechanism), that except for $^{136}$Xe, the other four nuclei have similar overall $0\nu\beta\beta$ NME's of values about $3$ ( see table \ref{lnt}) without quenching, when quenching is included, these values decrease to something between $2$ and $2.5$ (see table \ref{lnt}). For $^{136}$Xe, previous spherical calculation shows that it is relatively small due to the magicity of the neutron number, and the current calculation further reduces its value, and it is now only $1/3$ of the value of the other nuclei. This smallness of $^{136}$Xe was also observed in Skyrme QRPA calculations of \cite{ME13}, but in their calculation, they also got very small NME for $^{130}$Te, which is not observed in our calculation.

\begin{table*}[htp]
\caption{$0\nu\beta\beta$ matrix elements as in Table.\ref{lnt} but for heavy neutrino mechanism.}
\begin{center}
\renewcommand{\arraystretch}{1.5}
\begin{tabular}{|c|c|cccc|cccc|cccc|cccc|}
\toprule
& &\multicolumn{8}{c|}{Argonne}&\multicolumn{8}{c|}{CD Bonn}\\
\hline
& &\multicolumn{4}{c|}{$g_A=g_{A0}$}&\multicolumn{4}{c|}{$g_A=0.75g_{A0}$} 
&\multicolumn{4}{c|}{$g_A=g_{A0}$}&\multicolumn{4}{c|}{$g_A=0.75g_{A0}$} \\
& &  $M_{F,h}^{0\nu}$ &$M_{GT,h}^{0\nu}$& $M_{T,h}^{0\nu}$ & $M'^{0\nu}_h$
& $M_{F,h}^{0\nu}$ & $M_{GT,h}^{0\nu}$ & $M_{T,h}^{0\nu}$&$M'^{0\nu}_h$
&  $M_{F,h}^{0\nu}$ &$M_{GT,h}^{0\nu}$& $M_{T,h}^{0\nu}$ & $M'^{0\nu}_h$
&$M_{F,h}^{0\nu}$ & $M_{GT,h}^{0\nu}$ & $M_{T,h}^{0\nu}$&$M'^{0\nu}_h$ \\ 
  \hline
  $^{76}$Ge$\rightarrow ^{76}$Se 
  & a  &  -109.7 & 369.7 & -59.0 & 378.7& -109.5 & 423.1 & -63.5 &270.2
           &  -111.0  & 370.8 & -54.2 &385.4 & -110.8 & 426.6 & -58.2 &275.9\\
  & b & -83.2   & 198.0 & -62.2 &187.3 & -83.1   & 206.1 & -67.0 &129.7
            & -102.1 & 287.8  & -57.4 &293.7 &-101.9 &317.8 & -61.8 &207.2\\
  \hline
  $^{82}$Se$\rightarrow ^{82}$Kr 
  & a  & -102.3 & 345.9 & -54.1  & 355.3 & -102.2 & 397.0 & -58.0& 254.1
         & -102.5  & 344.2 & -48.9 & 358.7 & -102.4 & 397.1 & -52.4& 257.4\\
  &  b &  -77.4   & 184.9 & -57.0 & 175.9 &-77.3   & 193.2 & -61.2& 122.1
         & -94.2    & 267.0  & -51.8 & 273.6 & -94.1 &295.8 & -55.6 & 193.4\\
  \hline
   $^{130}$Te$\rightarrow ^{130}$Xe
  & a &  -116.1 & 393.0 & -68.8 & 396.3& -116.0 & 440.6 & -74.5 &277.9
      & -115.9  & 391.5 & -62.3 &401.1 & -115.9 & 439.7 & -67.5 &281.2\\
  & b &  -87.7   & 209.5 & -72.5 &191.4 &-87.6   & 213.6 & -78.7 &130.2
      & -106.4 & 303.4  & -65.9 & 303.5& -106.4 &326.9 & -62.5 &209.5\\
  \hline
  $^{136}$Xe$\rightarrow ^{136}$Ba 
  & a & -37.8 & 133.6 & -21.8 &135.2& -37.8 & 153.2 & -23.2&96.5
      & -32.5  & 113.0 & -16.1 &117.1& -32.4 & 128.5 & -17.1&82.7\\
  & b  &  -28.8   & 72.0 & -23.0 &66.9&-28.8   & 75.0 & -24.5&46.3
      & -30.2& 88.7  & -16.9 &90.5& -30.2 &96.5 & -18.1&62.8\\
%  \hline
%  $^{136}$Xe$\rightarrow ^{136}$Ba 
%  & a & -123.0 & 399.1 & -67.3 && -123.0 & 444.8 & -73.3&
%      & -125.4  & 404.3 & -57.7 && -125.4 & 447.3 & -62.6&\\
%  & b  &  -92.8   & 212.1 & -71.0 &&-92.8   & 215.0 & 77.4&
%      & -116.0& 315.4  & -60.8 && -116.0 &334.2 & -66.2&\\
  \hline
  $^{150}$Nd$\rightarrow ^{150}$Sm 
  & a & -127.4 & 414.7 & -70.2& 423.5 & -127.4 & 466.8 & -75.3& 299.2
        & -130.8  & 420.0 & -59.0& 442.0 & { -127.8} & { 466.1} & { -69.2}& 302.5\\
  & b & -96.3   & 220.4 & -74.0 & 206.1 & -96.2  & 226.5 & -79.5& 142.3
         & -120.1 & 325.0  & -62.4 & 337.0 & { -117.3} & { 346.6} & { -73.4}&226.4\\
  \hline
\botrule
\end{tabular}
\end{center}
\vspace{-0.5cm}
\label{hnt}
\end{table*}

%heavy neutrino
With the normal hierarchy (NH) the light neutrino mechanism (LNM) can be suppressed due to a small $0\nu\beta\beta$ effective Majorana neutrino mass. In this case one expects, that the heavy neutrino mechanism (HNM) becomes dominant. Therefore, we also calculate the $0\nu\beta\beta$ NME's for this mechanism and present the results in Table \ref{hnt}. Like for the light neutrino mechanism  (LNM), if src is not considered, different NN forces (AV18 and CD-Bonn)  behaves similarly, the difference of the NME's is small, for most nuclei, the deviations are under $1\%$.

 The $|\chi_F|$ are now mostly close 1/3, if quenching is not included. This value is very close to the shell model value, but if quenching is considered, this value will reduce to about $1/4$. Tensor contributes about $15\%$ to the reduction of the  axial vector current vertices ($J^\mu_A$). The NME's with the HNM have similar dependence on $g_{pp}$'s as for the LNM. The inclusion of quenching is changing the Fermi NME's by only a few percent probably due to numerical errors. $M_{GT,h}^{0\nu}$ has been enhanced by $10\%$, this is much smaller than the $30\%$ enhancement of $M_{GT,l}^{0\nu}$ for LNM, this implies that HNM NME's are less sensitive to $g_{pp}$'s. Therefore, the overall NME's for HNM are more reduced by quenching while $M^{0\nu}_{T,h}$ is insensitive to $g_{pp}$. For the nuclei calculated we find that this reduction for the HNM could be as high as $30\%\sim40\%$, significantly larger than that for LNM. Therefore, understanding the origin of quenching and obtaining  its exact value could be of great importance towards a better prediction of NME's. The realistic NME's are generally around 400 (see  table \ref{hnt}) without quenching and reduced to below 300 when quenching is included except for $^{136}$Xe, which is only about 1/3 to common NME values.

From Table \ref{hnt}, we see that now src plays a much more important role for HNM. This could be easily understood, because under LNM  neutrino is light, therefore the effective interaction mediated by neutrino behaves like a long range Coulomb interaction, the interplay between nucleons at short distance are not so important. But for HNM, because of the large neutrino masses, the neutrino produces an effective point-like interaction. Now the strong repulsive nature of nucleon-nucleon forces  at short distance must be seriously considered. Unlike for LNM, for HNM the reduction of src for NME's can be as high as $20\%$ for the Fermi part and nearly $50\%$ for GT part. But the effect of src to Tensor is much smaller. Argonne src's are much stronger than CD-Bonn src's as before. But even the milder CD-Bonn src's give an about $25\%$ reduction to GT. The correlations at short distance are due to a hard repulsive core of these two NN interactions, since the src's are obtained consistently from the potential. Due to the strong repulsive nature of nuclear force at short distance, we now have a reduction of about $50\%$ for Argonne src and $25-30\%$ for CD-Bonn src for $M^{0\nu}_{GT,h}$. Now with AV18 for all studied nuclei except for $^{136}$Xe, the total $0\nu\beta\beta$ NME's are around 200 without quenching and below $150$ with quenching. For $^{136}$Xe, these values are reduced to about 60 and below 50 respectively. For the CD-Bonn interaction, the common overall NME's are about 300 (see table \ref{hnt}) for cases without quenching and 200 with quenching, again $^{136}$Xe has NME's of about 1/3 of the other  above nuclei. The deviation due to the choice of src's now increases to about $25\%$ of the standard NME's without src's.
These values of NME's for the two mechanisms may play a very important role for the determination of the neutrino masses once $0\nu\beta\beta$ is observed.

\begin{table}[htp]
\caption{Comparison of $0\nu\beta\beta$ NME's  from different approaches for both light and heavy neutrino mass mechanisms. Here the parameters for this work and QRPA-T\"u\cite{SRFV13,FGK14}, IBM-2\cite{BKI13} and CDFT\cite{SYR14} are unquenched $g_A$ and Argonne src. QRPA-Jy\cite{HS15} uses CD-Bonn src and QRPA-NC\cite{ME13} is without src and tensor part contributions.}
\begin{center}
\begin{tabular}{|c|c|ccccc|}
\hline
 				& methods & $^{76}$Ge & $^{82}$Se & $^{130}$Te & $^{136}$Xe &$^{150}$Nd \\
\hline
\multirow{5}{0.8cm}{LNM} &this work 				& 3.12 & 2.86 & 2.90 & 1.11 & 3.01\\
					& QRPA-T\"u\cite{SRFV13} 	& 5.16 & 4.64 & 3.89	& 2.18 & - \\
					& QRPA-Jy \cite{HS15} 		& 5.26 & 3.73 & 4.00 & 2.91 & - \\
					& QRPA-NC\cite{ME13} 		& 5.09 & 	-	& 1.37 & 1.55 & 2.71\\
					& IBM-2 \cite{BKI13} 		& 4.68 & 3.73 & 3.70 & 3.05 & 2.67 \\
					& CDFT\cite{SYR14} 		& 6.04 & 5.30 & 4.89 & 4.24 & 5.46 \\
					& ISM\cite{Men17}			& 2.89 & 2.73 & 2.76 & 2.28 & - \\
\hline
\multirow{5}{0.8cm}{HNM} & this work				& 187.3 & 175.9 & 191.4 & 66.9 & 206.1 \\
					& QRPA-T\"u \cite{FGK14} 	& 287.0 & 262.0 & 264.0 & 152.0 & -	\\
					&  QRPA-Jy\cite{HS15}		& 401.3 & 287.1 & 338.3 & 186.3 & - \\
					& IBM-2\cite{BKI13} 			& 104 & 82.9 & 91.8 & 72.6 & 116 \\
					&  CDFT\cite{SYR14}  		&209.1 & 189.3 & 193.8 & 166.3 & 218.2 \\
					& ISM\cite{Men17}			& 130  & 121	 & 146   &  116 & -      \\
\hline
\end{tabular}
\end{center}
\label{dfm}
\end{table}

%Comparison of different approaches
Various approaches as stated in the section of "Introduction" have been adopted for the calculations of $0\nu\beta\beta $ NMEs, but up to now, deviations are still large among different methods, as is presented in Table.\ref{dfm}. For LNM, QRPA-T\"u\cite{SRFV13} and QRPA-Jy\cite{HS15} are without deformation, they differs only slightly for most nuclei but a larger deviation is observed for $^{136}$Xe (Most methods uses Argonne src except QRPA-Jy(CD-Bonn) and QRPA-NC(no src)). The deviation of this work to these two QRPA calculations with realistic force is explained above. For QRPA-NC\cite{ME13}, one got close NMEs to this work for $^{136}$Xe and $^{150}$Nd but large deviations for $^{76}$Ge and $^{130}$Te. For approaches with closure approximation, IBM-2\cite{BKI13} has results close to the QRPA calculations without deformation while CDFT method\cite{SYR14} obtained much larger NMEs compared to all other methods, and for ISM\cite{Men17}, we got close results for most nuclei except $^{136}$Xe. For HNM, the situation becomes a bit different, the deviation between this work and the two other QRPA methods with realistic force can be explained as the deformation effects, and the difference between QRPA-T\"u and QRPA-Jy are solely due to the different src's(Their comparison for the same src can be found in\cite{HS15}). For most nuclei, IBM-2 method has now NMEs way smaller than any other methods. The same happens for ISM, it has the second smallest NMEs among all the methods, and compared to this work, their results are generally smaller except $^{136}$Xe. The NMEs from CDFT seems to agree with current work for most nuclei except $^{136}$Xe. The origin of deviations of these approaches still needs investigation and we need to sort out how much the closure approximation would alter the final results.
 In general, we still need to improve the accuracy of nuclear many-body calculations for a better predictions of decay rates of these processes.
\section{Decomposition of $0\nu\beta\beta$ NME for $^{150}$Nd}
In this section, we will discuss the general structure of nuclear matrix elements for $^{150}$Nd, how different parts contribute to the final results. We also compare results from different model spaces to understand how model space truncations affects the final NME's. For this purpose, we use three different model spaces: I for N=4-6 (3 major shells), II for N=0-6 (7 major shells) and III for N=0-7 (8 major shells), respectively (hereafter called MSp-I, MSp-II and MSp-III). The MSp-I includes only the shells containing the Fermi-levels of either neutrons or protons and their neighboring shells. These are the most active shells, which are only partially occupied and are close to the Fermi level with smaller quasi-particle energies. The MSp-II adds all the nearly occupied shells below MSp-I.  MSp-III contains also nearly unoccupied shells above MSp-I besides the shells of MSp-II. Such different  model spaces will help to understand how the partially-occupied shells, the occupied ones and the non-occupied ones contribute to the NME's. 

As in most QRPA calculations, we fix the parameters $g_{pp}$'s by reproducing the experimental $2\nu\beta\beta$ nuclear matrix elements. For our investigation of the structure of the $0\nu\beta\beta$ NME's, we use parameters as follows: quenched $g_A=0.75g_{A0}$, CD-Bonn potential with CD-Bonn src's (short range correlations). The fitted $g_{pp}$ values are: $g_{pp}^{T=0}=1.03$, $0.78$, $0.76$ for model space I, II, III and $g_{pp}^{T=1}=1.34$, $1.09$ and $1.06$, respectively. There are large changes of $g_{pp}$'s of more than $30\%$  from the severely truncated MSp- I to II and III, since the $2\nu\beta\beta$ NME's decrease drastically with increasing  particle-particle strength \cite{YRF08}. The parameters $g_{pp}$'s are fitted to larger values to obtain agreement for the $2\nu\beta\beta$ transition probabilities for a smaller model space. The values of $g_{pp}$ for model space MSp-II and MSp-III are very close. This implies, that for the $2\nu\beta\beta$ decay, orbits inside 7 major shells of space MSp-II N=0-6 give most of contributions to the NME's. Results with more major shells do not change the matrix elements appreciably.

In the above parametrization for the truncated model spaces larger particle-particle strengths  have been used to compensate the reduction of the NME's due to the reduced model space. With correctly reproduced $2\nu\beta\beta$ NME's, this means that for $2\nu\beta\beta$ the compensation from increased strength parameters $g_{pp}$'s equals to the reduction from excluded shells missing in truncated model space calculations. So for $0\nu\beta\beta$ calculations with truncated model space, the errors comes from a competition of above compensation and reduction (or enhancement since $0\nu\beta\beta$ may behave differently 
than $2\nu\beta\beta$).

\begin{figure*}
\includegraphics[scale=0.5]{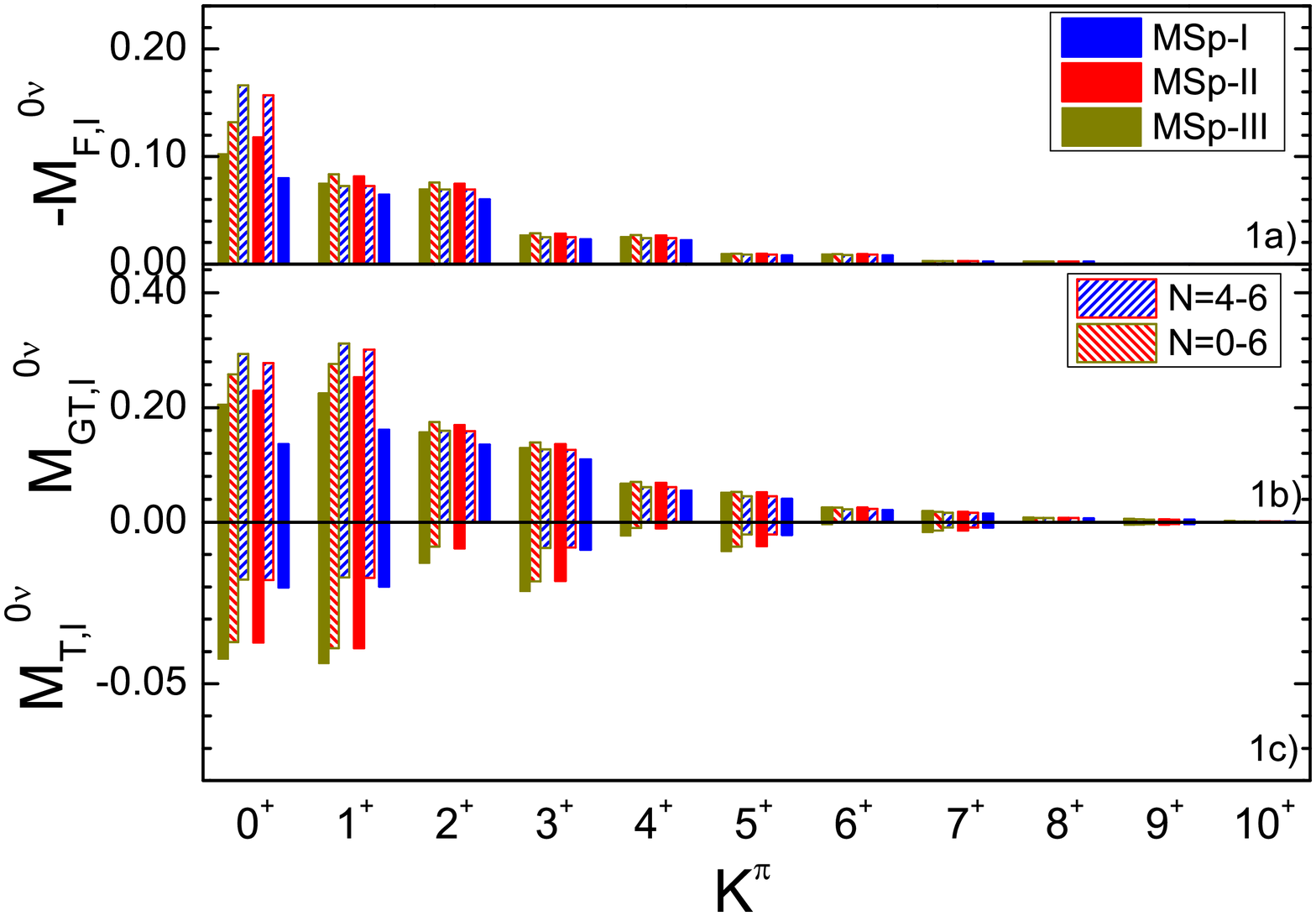}
\includegraphics[scale=0.5]{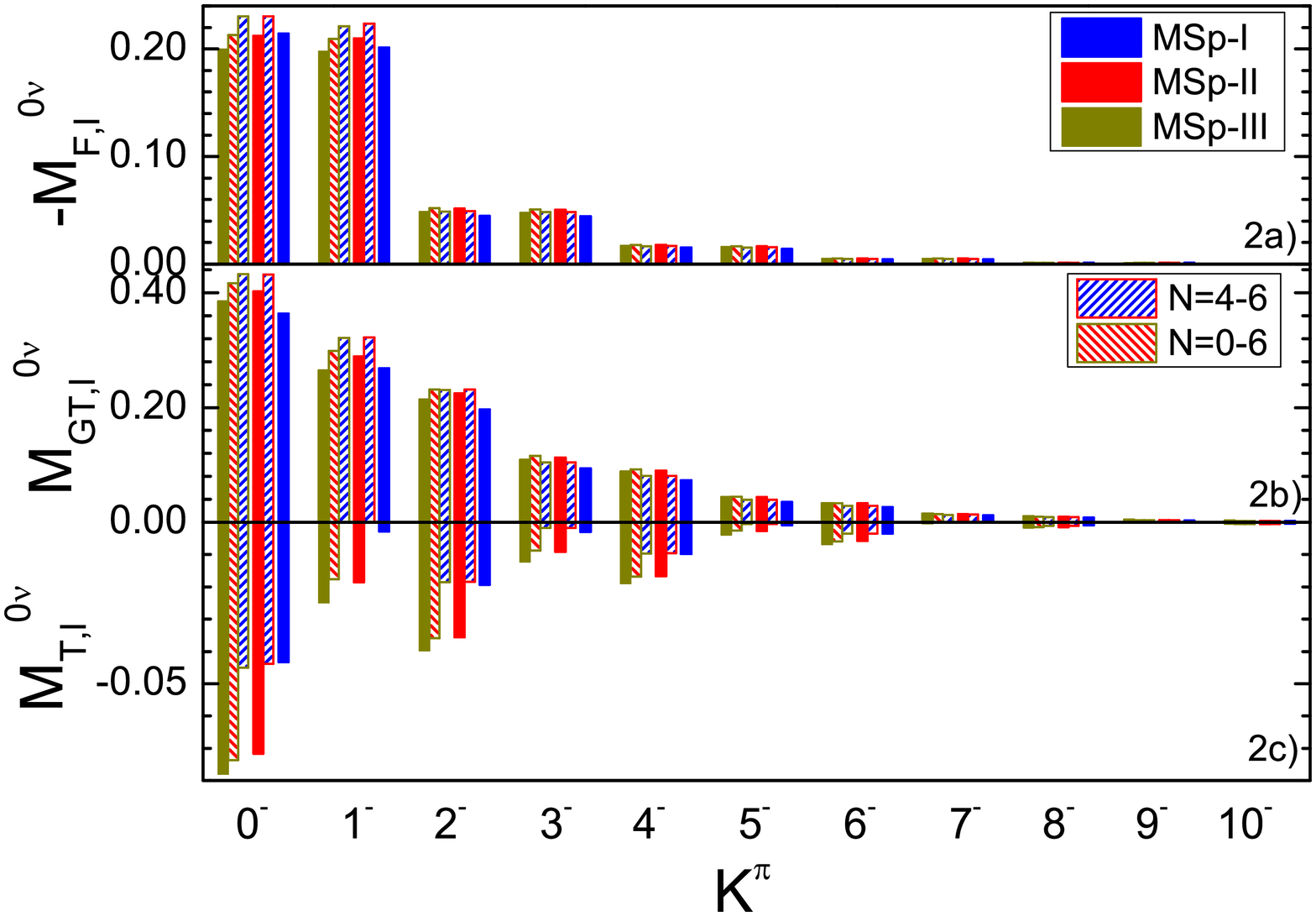}
\caption{(Color online) Decomposition of the matrix elements ($M_F^{0\nu}$, $M_{GT}^{0\nu}$ and $M_T^{0\nu}$) over $K^\pi$ intermediate states for positive (upper panel) and negative (lower panel) parities for $^{150}$Nd. The solid bars for a fixed $K^{\pi}$ correspond from left to right are MSp-III for N = 0-7 (8 major shells), MSp-II for N = 0-6 (6 major shells) and MSp-I for N = 4-6 (3 major shells). The different colors correspond to different model spaces as indicated in the figures, the shaded areas to the right of the solid bars are contributions from sub spaces in larger space with their size indicated in the figure.}
\label{lnh}
\end{figure*}

\begin{figure*}
\includegraphics[scale=0.5]{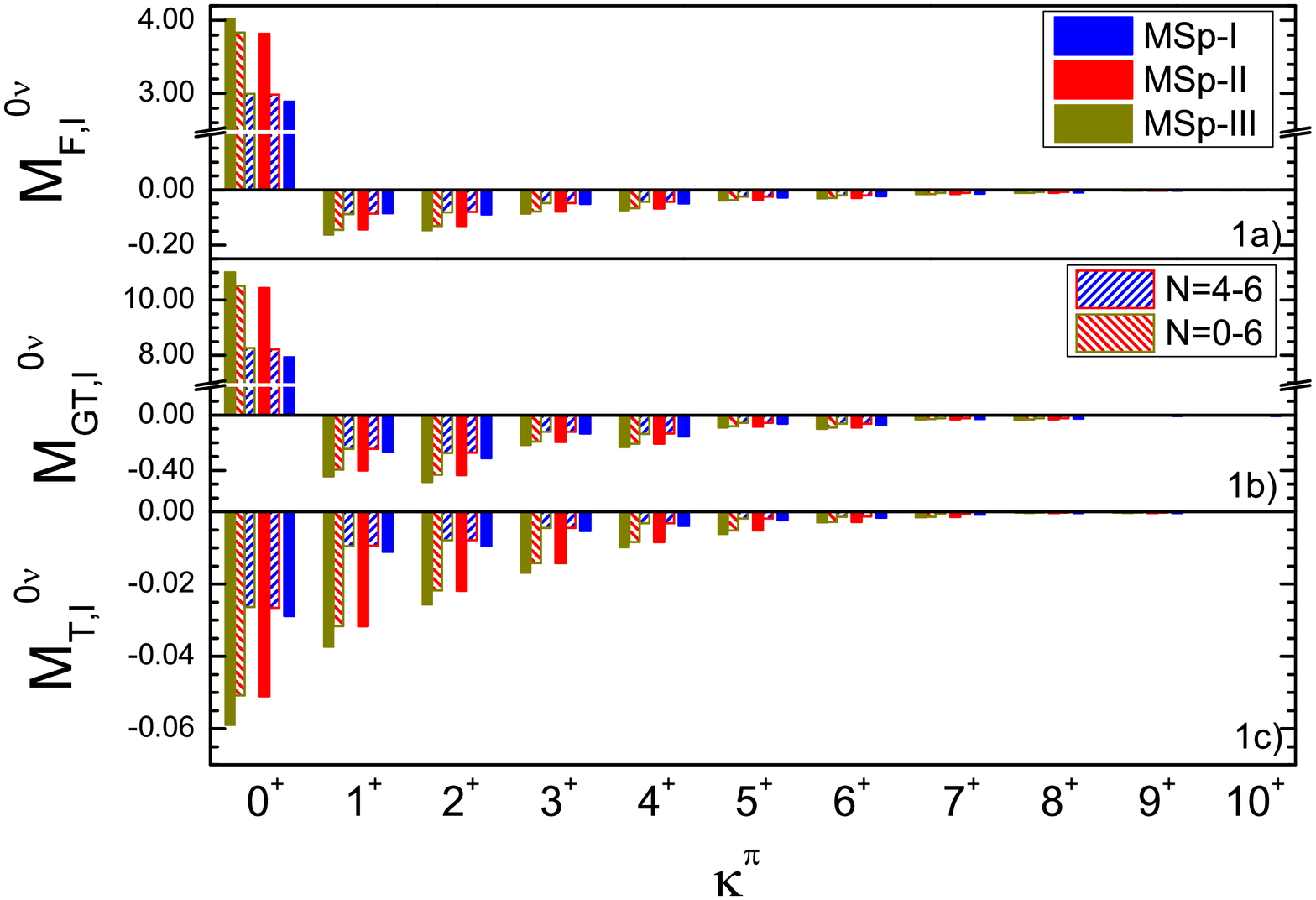}
\includegraphics[scale=0.5]{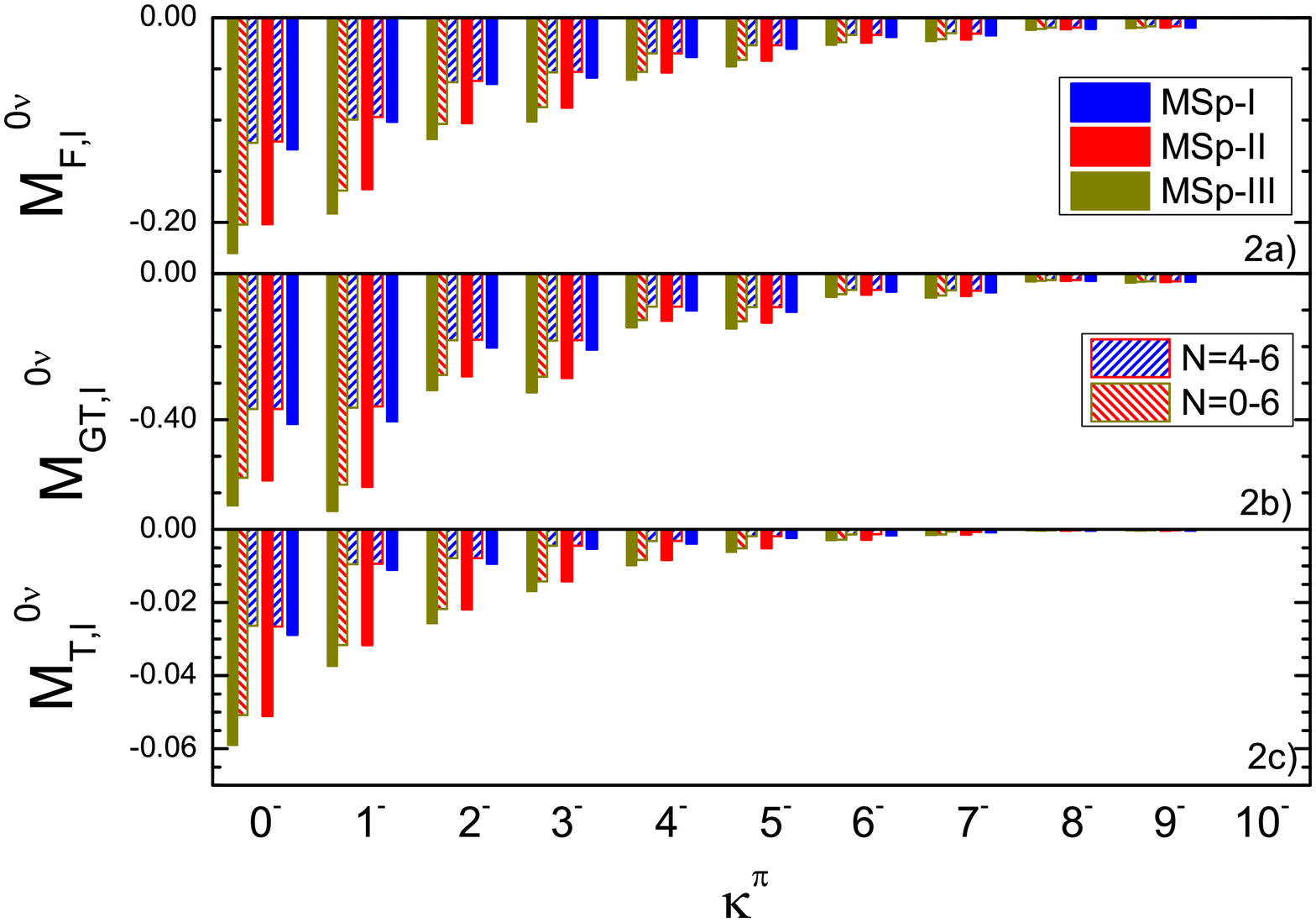}
\caption{(Color online) Decomposition of the matrix elements ($M_F^{0\nu}$, $M_{GT}^{0\nu}$ and $M_T^{0\nu}$) over different $\kappa^\pi$ of decaying nucleon pairs for positive (upper panel) and negative (lower panel) parities for $^{150}$Nd. The solid bars for a fixed $\kappa^{\pi}$ correspond from left to right are MSp-III for N = 0-7 (8 major shells), MSp-II for N = 0-6 (6 major shells) and MSp-I for N = 4-6 (3 major shells). The different colors correspond to different model spaces as indicated in the figures, the shaded area on the right of the solid bars are contributions from sub spaces in larger space as indicated in the figure.}
\label{lnp}
\end{figure*}

\begin{figure*}
\includegraphics[scale=0.5]{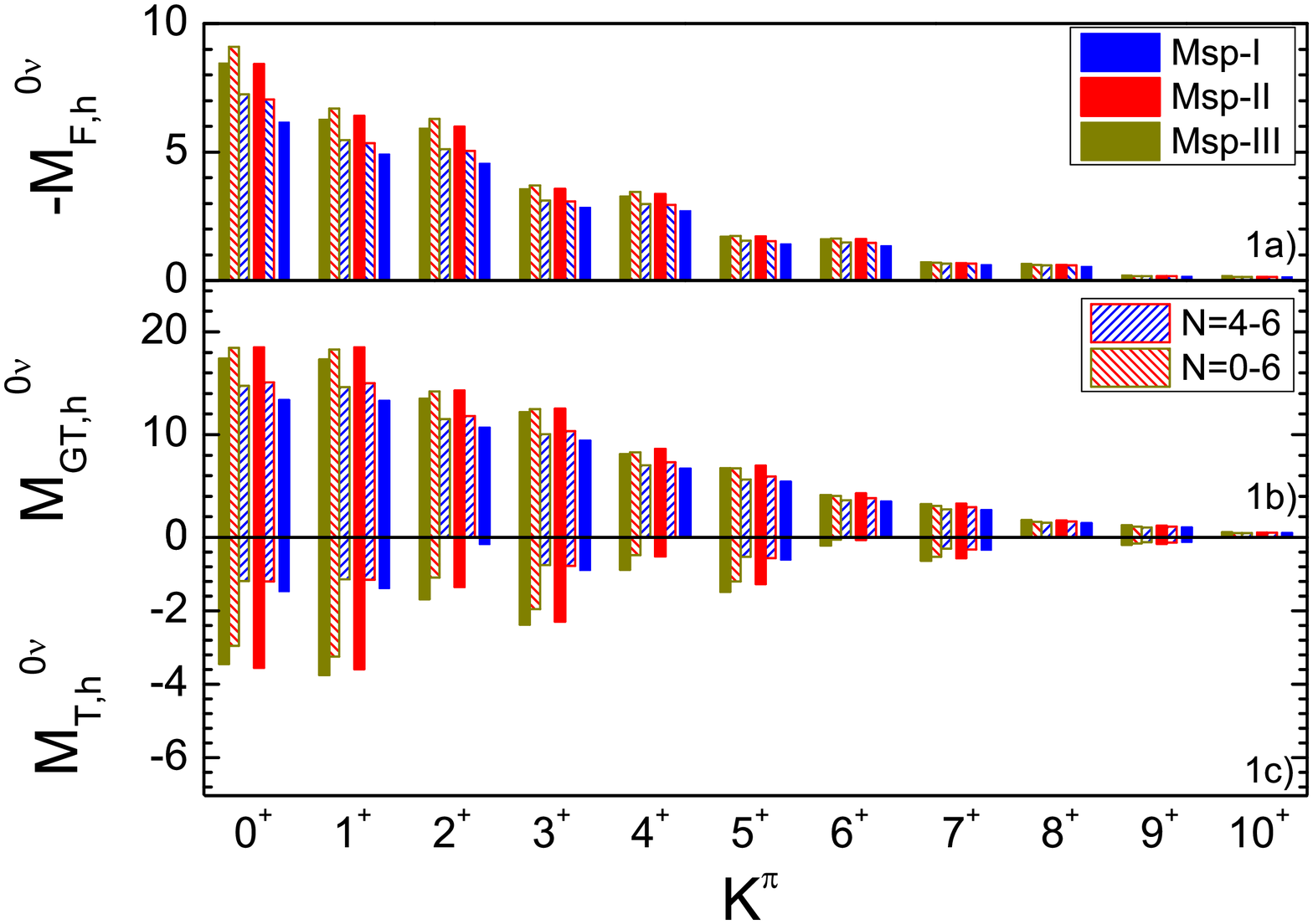}
\includegraphics[scale=0.5]{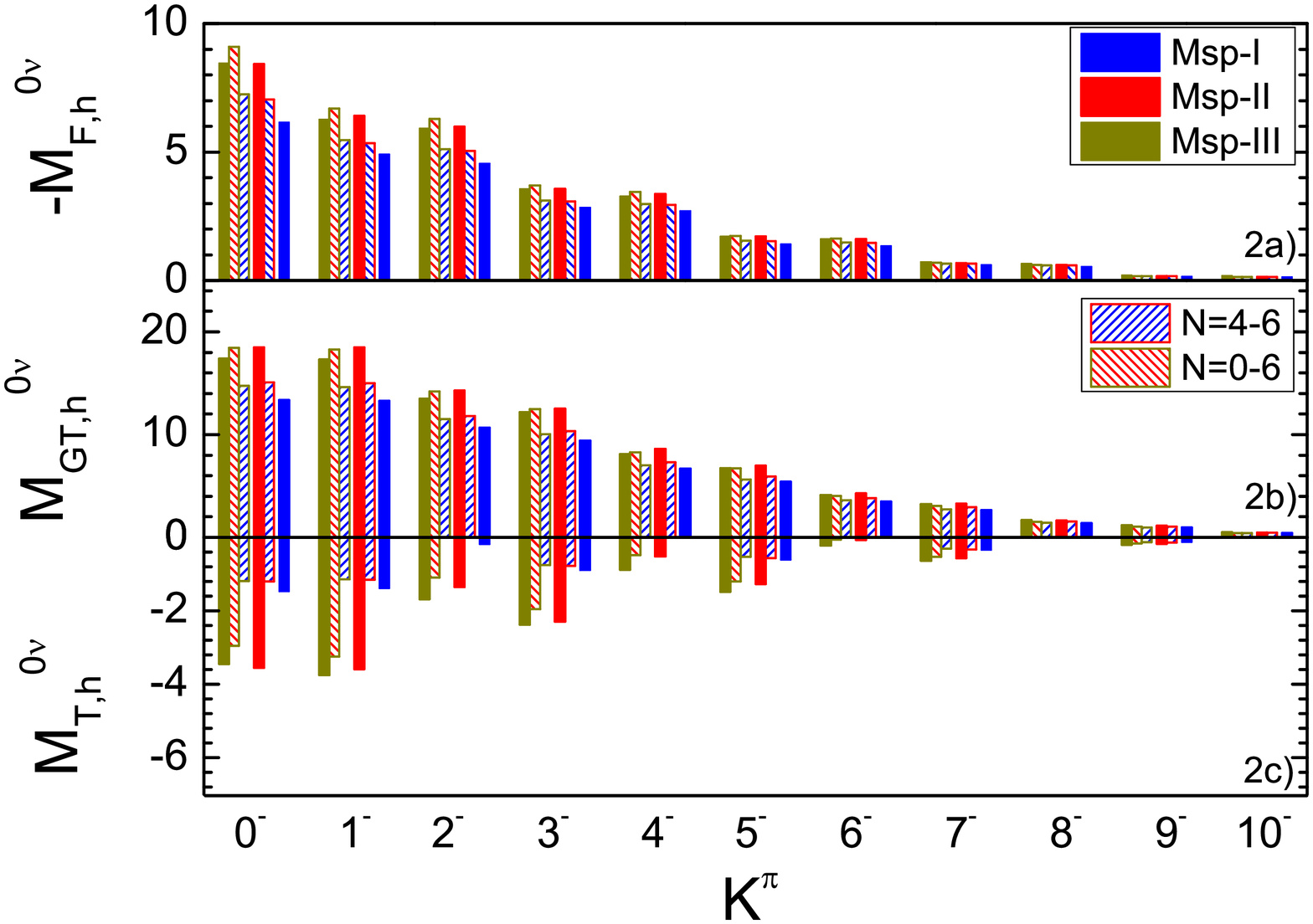}
\caption{(Color online) The same as Fig.\ref{lnh} but for the heavy neutrino mechanism.}
\label{hnh}
\end{figure*}

\begin{figure*}
\includegraphics[scale=0.5]{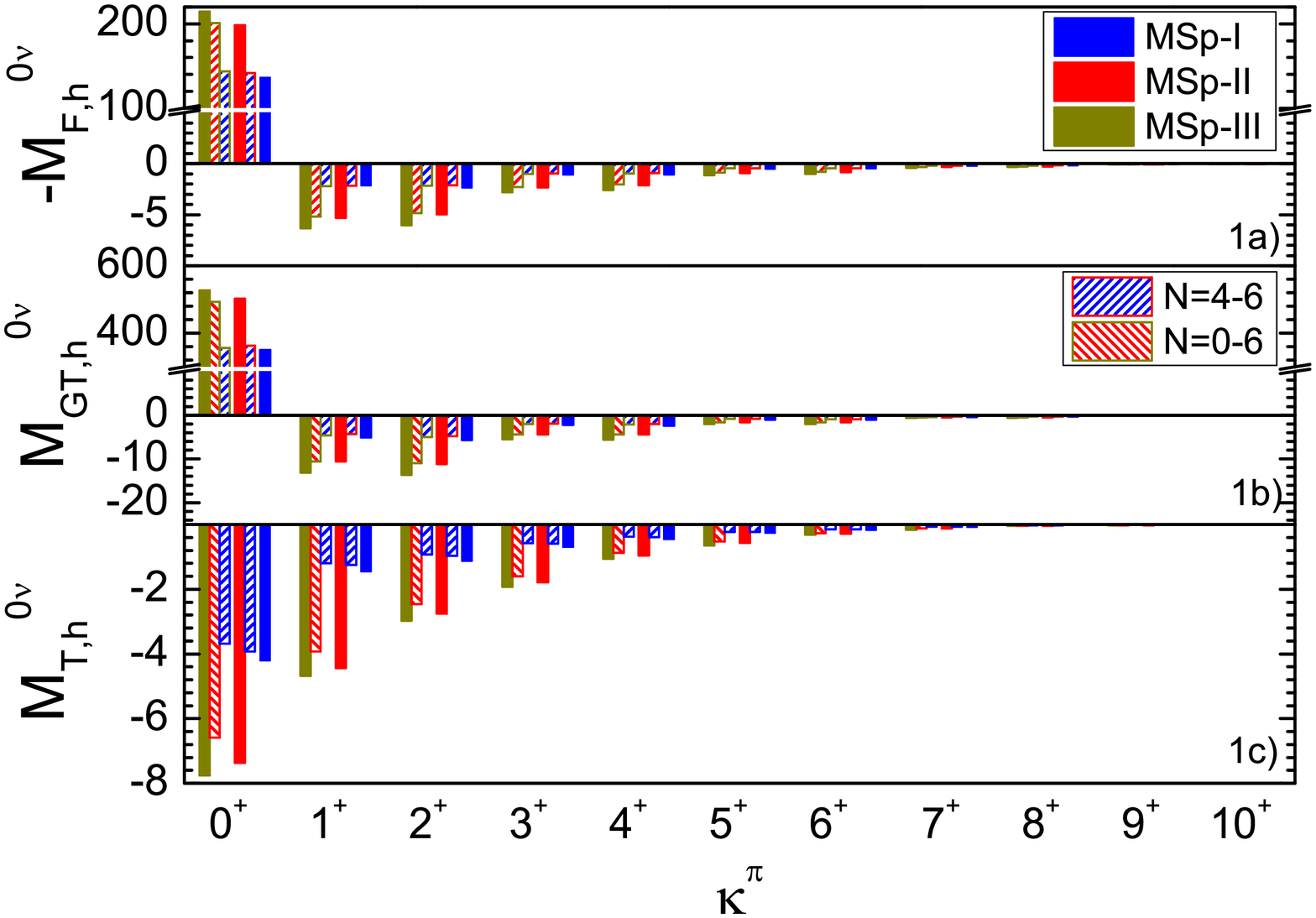}
\includegraphics[scale=0.5]{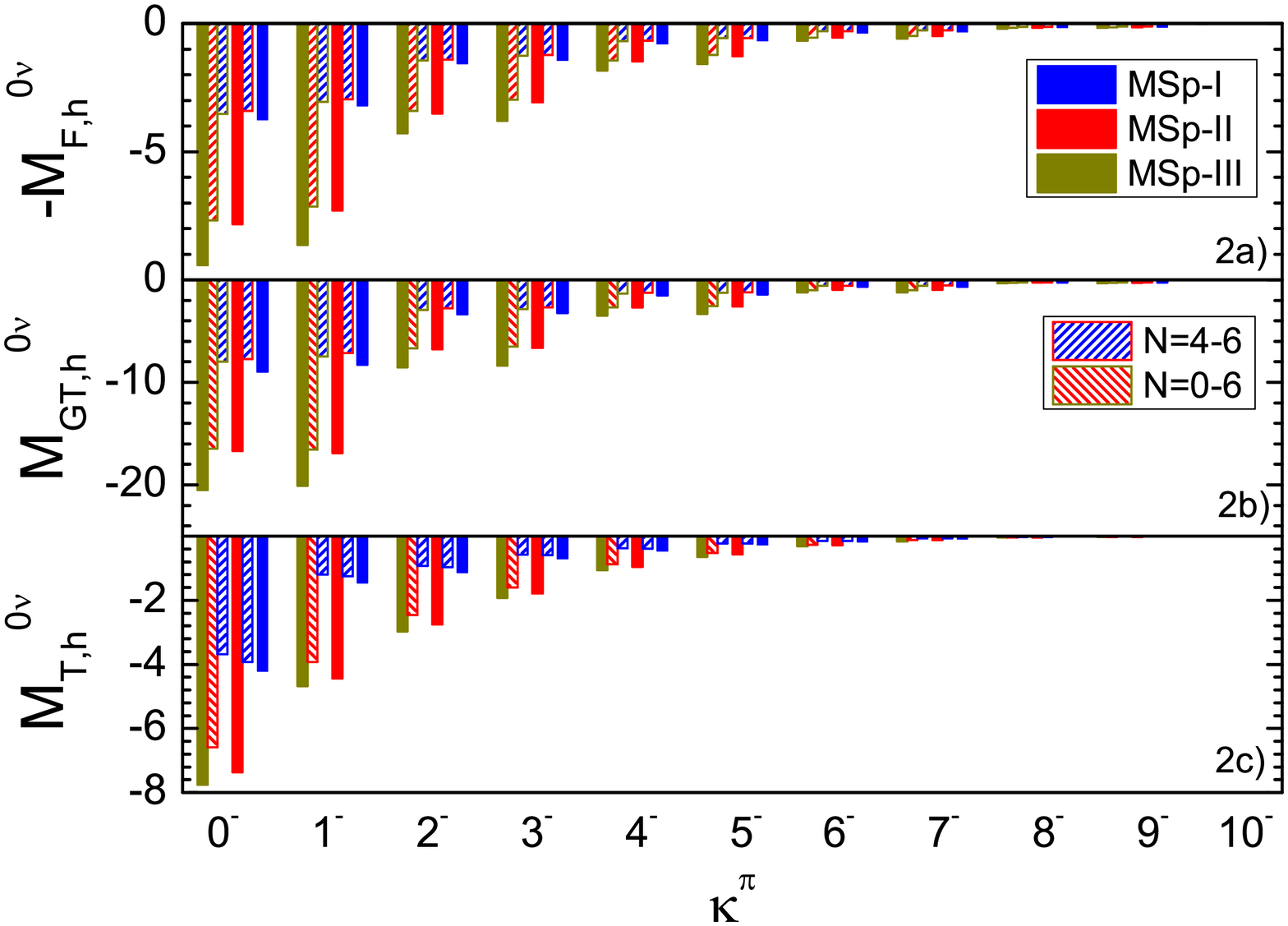}
\caption{(Color online) The same as Fig.\ref{lnp} but for the heavy neutrino mechanism.}
\label{hnp}
\end{figure*}

Figs.\ref{lnh}, \ref{lnp}, \ref{hnh} and \ref{hnp} help to understand this behavior: The three solid bars of each $K^\pi$ correspond to overall results for the three model spaces with size from large to small (MSp-III for N=0-7, 8 major shells; MSp-II for for N=0-6, 7 major shells and MSp-I for N=4-6, 3 major shells, respectively), when going from left to right. In a larger model space, contributions from a smaller sub-space (the same color denoted the same size of the space or sub-space) are plotted by the shaded bars. The differences between the solid and shaded bars are the contributions from the parts where excluded shells are participating. For specific $K^\pi$, the bars with the same color could also present results with different $g_{pp}$'s with a increasing order from left to right. As we discussed above, for different model spaces we have different $g_{pp}$'s. Thus the first MSp-III and the second bar MSp-II have very close values of $g_{pp}$ and there is a drastic difference between the first two blue shaded bars and the third solid bar.

The restoration of isospin symmetry leads to a reduction of only $K^\pi=0^+$ contributions of $M_F^{0\nu}$. This has been discussed in \cite{FFS15}, and is not the topic of the current article. In this work we have also included the Tensor part from the induced hadronic current. Its general effect has been discussed in the previous section. In this section, we will discuss the partial contributions from different $K^\pi$'s below. 

Once again, we start with the discussion of the light neutrino case (fig.\ref{lnh}), considering how the different intermediate states contribute. Calculations in the three different model spaces are parametrized to reproduce exactly the same experimental $2\nu\beta\beta$ NME's. But the results for $0\nu\beta\beta$ are obviously not any more the same. 
%As explained above, for $2\nu\beta\beta$ with a smaller space, larger $g_{pp}$'s or namely over-correlations in particle-particle interactions are needed to compensate the necessary reductions from transitions not included. This over-correlation in a truncated model space doesn't somehow compensate the $2\nu\beta\beta$ and $0\nu\beta\beta$ simultaneously.  

\begin{table}[htp]
\caption{$0\nu\beta\beta$ NME's of $^{150}$Nd for different Model spaces and contributions from the sub-spaces.}
\begin{center}
\begin{tabular}{|c|ccc|cc|c|}
\hline
 				& \multicolumn{3}{c|}{MSp-III} & \multicolumn{2}{c|}{MSp-II}& MSp-I \\
				& Full 	& N=0-6 	& N=4-6 	& Full 	& N=4-6 	& Full \\
\hline
$M_{F,l}^{0\nu}$	& -1.43	& -1.55	& -1.55	& -1.52	& -1.55	& -1.34 \\
$M_{GT,l}^{0\nu}$ 	& 3.55	& 3.91	& 3.91	& 3.75	& 3.87	& 3.04 \\
$M_{T,l}^{0\nu}$	& -0.53	& -0.44	& -0.18	& -0.44	& -0.18	& -0.21 \\
\hline
$M_{F,h}^{0\nu}$ 	& -120.1	& -124.9	& -109.3	& -120.2	& -107.7	& -98.8 \\ 
$M_{GT,h}^{0\nu}$ 	& 325.0	& 331.8	& 280.7	& 339.1	& 290.9	& 264.4 \\
$M_{T,h}^{0\nu}$	& -62.4	& -51.5	& -21.2	& -57.5	& -22.0	& -25.5  \\
\hline
\end{tabular}
\end{center}
\label{partr}
\end{table}

The general contributions from different $K^\pi$ are similar for $M_F^{0\nu}$ and $M_{GT}^{0\nu}$. The low K states contribute more, since low K states correspond to contributions near the Fermi surface. This partially holds also for the tensor part $M_{T}^0\nu$. Unlike for the GT part, where the NME's decrease nearly monotonically with increasing K, we see a staggering behavior for the size  of the tensor parts. We find a reductions for the so-called natural parity $\pi=(-1)^K$ states. Their contributions are much smaller than the unnatural ones. For all these parts of NME's of $^{150}$Nd, we find, that negative parity states are generally larger than positive states with similar K. This is somehow not the general case for all nuclei \cite{FFR11}, but for heavier nuclei only. Since heavier ones are generally neutron-richer, the proton and neutron Fermi levels are in different major shells with opposite parities. This makes the transitions with negative parity more favored and contributes more to $0\nu\beta\beta$ NME's.

%pp dependence more quantified
The $g_{pp}$ dependence for different $K^\pi$'s are different \cite{Fan11}, this can also be seen by comparing the blue solid bars and shaded blue areas. We should be aware, that the dependence on $g_{pp}$'s is different for the different matrix elements  \cite{SRFV13}: $M_F$ depends only on the isovector and $M_{GT}$ only on the isoscalar strength. Hereafter, we discuss only the corresponding $g_{pp}$ dependence. From the figures, we see, that for the Fermi part only $0^+$ and for the GT part only $0^+,\pm1^+$ are sensitive to $g_{pp}$, while for the tensor part, compared with GT or F, the results are basically $g_{pp}$ independent.  
%model space dependence   
Now we discuss the contributions of the different shells. For the step MSp-I to MSp-II one adds to the partially occupied orbits the occupied orbits (comparing the blues and reds in the first and second bars). We see reductions for several $K^\pi$'s mainly for low K's. The  NME's are enhanced for other multipoles (mainly high K's). To see their overall effects, we present these partial NME's in table \ref{partr}. The results show, that the high K enhancement is nearly equal to the low K reduction from the occupied levels. Therefore, when occupied shells are included, although states with different K behaves differently, the total NME's for GT and Fermi doesn't change too much, for them the increase of the particle-particle strength from fitting $2\nu\beta\beta$ NME's will compensate the reduction from model space truncation for the final results. The Tensor part behaves differently, for them the addition of the occupied shells reduces the overall NME's. The effect of the inclusion of unoccupied shells can be obtained by comparing the yellow bars and red shadows, for nearly all multipoles, these unoccupied orbits reduce the NME's. For GT and Fermi parts, these orbits reduces about $10\%$ of total NME's, and for Tensor parts, we have an additional reductions but much smaller. 
%And the same is also true when we add the orbits above MSp-I(red bars), the more shells further reduce the results and cancels the increasing trends of NME with decreasing $g_{pp}$. These are results for GT and Fermi part, for tensor part the effect is unique, that the magnitude of the NME increase as the model space being enlarged, since tensor part gives also negative values, the enlargement of model space will reduce the overall NME.

 Fig.\ref{lnh} and Table.\ref{partr} show that the roles of truncations for the final results are dependent on the choice of the specific truncations. For the  Light Neutrino exchange Mechanism (LNM) truncation of occupied levels tend to reduce the total NME's,  while the removal of unoccupied orbitals tends to enhance the overall NME's. Therefore, we can estimate how errors are generated in MSp-I and MSp-II by the results of model space MSp-III. The actual $g_{pp}$ in an infinite Hilbert space would be close to those fitted for MSp-III.  Since the unoccupied levels newly added in  MSp-III are far from the Fermi surface, they can  not  have a large impact to $2\nu\beta\beta$ NME's. These omitted levels are nearly unoccupied, so from the above analysis the removal of these levels from the model space will enhance the NME's. But the change of the NME's due to the step from MSp-II to MSp-III is much smaller than the difference of the results  between MSp-I and MSp-II. This implies that the errors of NME's with current model space should be much less than $10\%$.
%With the tiny change of $g_{pp}$, the change of $0\nu\beta\beta$ NME will mostly due to the model space change. Therefore a reasonable guess is that the errors should be much less than that of the difference between the results of MSp-II and MSp-III of a several percent.

How do the different nucleon pairs in the initial and final nuclei contribute to the NME's? In this work, we consider only the case $0\nu\beta\beta$ to ground states(namely $0^+\rightarrow 0^+$), this means that the total angular momentum projection $\kappa^\pi$ of decaying neutron pairs and total angular momentum $\kappa'^{\pi'}$ of rest nucleons in the nucleus should obey the relation: $\kappa+\kappa'=0$ and $\pi\cdot \pi'=1$, and the neutron pairs will decay to the proton pairs with the same $\kappa^\pi$ due to the conservation of angular momentum projections in the intrinsic frame. In the spherical case in QRPA \cite{EFR10} or Shell Model \cite{BHS14} calculations, one finds the dominant contributions from $J^\pi=0^+$ pairs to $0\nu\beta\beta$ NME's. This is also the case in our calculation (here this dominance translated to the dominance of $\kappa^\pi=0^+$ in the intrinsic system). We see that $0^+$ pairs consist of approximately three times the final NME for $M^{0\nu}_{F(GT)}$. This is due to the strong pairing correlations in the initial and final ground states as shown in Shell Model calculations \cite{BHS14}. Except $0^+$ pairs, all other pairs give reductions to the NME, and it is not surprising that lower $\kappa$'s give larger reduction since they are made of nucleons with orbits closer to each other. The pairs with negative parities are in different shells, they also give considerable reductions, this is due to the fact that the Fermi levels of neutron and proton are in different shells. 

For the truncation issues, we find that a smaller model space reduces the NME in the $0^+$ channel. This is reasonable, since for a truncated space fewer nucleon pairs are participating in the decay process. On the other hand reductions in other channels are weakened. The overall change of the NME's depends on the competition between the changes of the contributions  from $0^+$ and other channels. For the tensor part pairs with different $\kappa^\pi$ reduce the NME's. These reductions are enhanced, when more orbits are included and they are not sensitive to $g_{pp}$'s as the total tensor NME. 

For HNM the effective neutrino interactions between the nucleons involved in the decay is similar to a zero range  interaction instead of a Coulomb-like interaction for the LNM. As discussed in the previous section, $0\nu\beta\beta$ NME's with HNM have a weaker dependence on $g_{pp}$'s for both the total NME's as well as the partial NME's $M^{0\nu}(K^\pi)$ (see Figs (\ref{lnh},\ref{lnp},\ref{hnh},\ref{hnp})).  
%dependence for heavy neutrino mechanism 

For the HNM the removal of orbitals has a  different effect. The addition of occupied orbitals below MSp-I will enhance the NME's for the HNM for all the multipoles $K^\pi$ instead of only high K as for the LNM. Hence we observe a $15\%$ reduction to $M^{0\nu}_F$ and $M^{0\nu}_{GT}$, when the lower occupied orbitals (from MSp-II to MSp-I)  are removed. The unoccupied orbitals above MSp-II play for HNM a similar role as for LNM: $M^{0\nu}_{F(GT)}$ will be overestimated by a few percent, if they are removed from the model space. The tensor part for the HNM follows the pattern in the LNM and its reduction to the final NME's increase as the model space is enlarged. Table \ref{partr} shows, that the errors from the removal of the nearly unoccupied orbitals for HNM will be small.
%Unlike the LNM, inner shells $N=0-3$ gives enhancement instead of reduction to the final Fermi and GT NME's, the reason of such difference is discussed in \cite{} of ph force dominance.  In contrast the outer major shells gives small reductions to the NME. So when adding the N=7 major shells to the calculation, both the LNM NME and HNM NME are reduced. This makes estimations of how the model space changes the NME a bit difficult, we are not sure whether adding more shells will increase or reduce the NME, but we are confident, the changes will be tiny, less than several percent.

The decomposition of NME's for HNM over the decaying pairs follows similar pattern as that for LNM. Here the $0^+$ dominance holds and it contributes nearly two times of total NME's to the final NME's, which is then get reduced by other $\kappa^\pi$. The tensor reductions are getting enhanced when more shells are involved.

\section{Conclusions}
In this work, we calculated the $0\nu\beta\beta$ NME's with the deformed QRPA method with realistic forces for $^{76}$Ge, $^{82}$Se, $^{130}$Te, $^{136}$Xe and $^{150}$Nd. Both light and heavy neutrino exchange mechanisms are considered. We restored partially isospin symmetry. Our calculations show reductions of the NME's are mostly due to the BCS overlap factors which ranges from $0.7$ to $0.4$  depending on the nucleus. Our results shows that NME for $^{136}$Xe is heavily suppressed due to the magicity of the neutrons, while other nuclei have similar NME's. We also give the decomposition of NME's in different intermediate channels and estimate errors due to the truncation of the model space. 

\begin{acknowledgments}
This work is supported by the National Natural Science Foundation of China under Grant No. 11505078 and 11647304. F\v{S} is supported by the VEGA Grant Agency of the Slovak Republic under Contract No. 1/0922/16, by the Slovak Research and Development Agency under Contract No. APVV-14-0524, RFBR Grant No. 16-02-01104, Underground laboratory LSM - Czech participation to European-level research infrastructure CZ.02.1.01/0.0/0.0/16 013/0001733. We would also thank the anonymous referee for careful reading and useful comments and suggestions.
\end{acknowledgments}

\end{document}